\documentclass[lettersize,journal]{IEEEtran}
\usepackage{amsmath,amsfonts}
\usepackage{algorithmic}
\usepackage{algorithm}
\usepackage{array}
\usepackage[caption=false,font=normalsize,labelfont=sf,textfont=sf]{subfig}
\usepackage{textcomp}
\usepackage{stfloats}
\usepackage[switch]{lineno}
\usepackage{amssymb} 
\usepackage{xcolor}  
\usepackage{url}
\usepackage{verbatim}
\usepackage{graphicx}
\usepackage{cite}
\usepackage{booktabs}
\usepackage[colorlinks=true,
linkcolor=blue,       
citecolor=blue,       
urlcolor=blue,        
filecolor=blue        
]{hyperref}

\begin{document}
\title{Active Movable-Element RIS Assisted Vehicular Semantic Communications: Modeling and Optimization}

\author{Maoxin~Ji,
	Qiong~Wu,~\IEEEmembership{Senior~Member,~IEEE},
	Jingbo~Zhang,
	Pingyi~Fan,~\IEEEmembership{Senior~Member,~IEEE},\\
	Kezhi~Wang,~\IEEEmembership{Senior~Member,~IEEE},
	Wen~Chen,~\IEEEmembership{Senior~Member,~IEEE},\\
	Guoqiang~Mao,~\IEEEmembership{Fellow,~IEEE},
	and~Khaled~B.~Letaief,~\IEEEmembership{Fellow,~IEEE}
\thanks{This work was supported in part by Jiangxi Province Science and Technology Development Programme under Grant 20242BCC32016; in part by the National Natural Science Foundation of China under Grant 61701197; in part by Basic Research Program of Jiangsu under Grant BK20252084; in part by the National Key Research and Development Program of China under Grant 2021YFA1000500(4); in part by the 111 Project under Grant B23008 and in part by the Hong Kong Research Grant Council under Grant No. 16215624 as well as the Area of Excellence (AoE) Scheme under Grant No. AoE/E-601/22-R. (Corresponding author: Qiong Wu.)
	
	Maoxin Ji, Qiong Wu and Jingbo Zhang are with the School of Internet of Things Engineering, Jiangnan University, Wuxi 214122, China, and also with the School of Information Engineering, Jiangxi Provincial Key Laboratory of Advanced Signal Processing and Intelligent Communications, Nanchang University, Nanchang 330031, China (e-mail:  maoxinji@stu.jiangnan.edu.cn;
	qiongwu@jiangnan.edu.cn; jingbozhang@stu.jiangnan.edu.cn).
	
	Pingyi Fan is with the Department of Electronic Engineering, State Key Laboratory of Space Network and Communications, and the Beijing National Research Center for Information Science and Technology, Tsinghua University, Beijing 100084, China (e-mail: fpy@tsinghua.edu.cn).
	
	Kezhi Wang is with the Department of Computer Science, Brunel University of London, London, Middlesex UB8 3PH, U.K (e-mail: Kezhi.Wang@brunel.ac.uk).
	
	Wen Chen is with the Department of Electronic Engineering, Shanghai Jiao
	Tong University, Shanghai 200240, China (e-mail: wenchen@sjtu.edu.cn).
	
	Guoqiang Mao is with the School of Transportation, Southeast University, Nanjing, China (email: g.mao@ieee.org).
	
	Khaled B. Letaief is with the Department of Electrical and Computer Engineering, the Hong Kong University of Science and Technology, Hong Kong (email: eekhaled@ust.hk).}}



\maketitle

\begin{abstract}
Severe signal blockage and fast-varying channels in vehicular environments pose critical challenges to reliable semantic communication. To address these, this paper proposes a novel Row-Movable Active Reconfigurable Intelligent Surface (RM-A-RIS) assisted vehicular semantic communication system. This architecture uniquely combines active signal amplification with element mobility to compensate for multiplicative fading and reconstruct channel geometry, thereby enhancing spatial diversity. We formulate a joint optimization problem to maximize Semantic Spectral Efficiency (SSE) by coordinating RIS element positions, active reflection coefficients, and semantic symbol length. An efficient Alternating Optimization (AO) algorithm is developed to tackle the coupled non-convexity. Simulation results demonstrate that the proposed scheme substantially outperforms existing benchmarks, achieving up to 132.9\%, 9.2\%, and 35.2\% improvements in Sum-Semantic Spectral Efficiency (Sum-SSE) compared to the passive RIS, fixed-position active RIS, and QPSO baselines, respectively.
\end{abstract}


\section{Introduction}
\IEEEPARstart{W}{ith} the evolution of 6G, the Internet of Vehicles (IoV) has emerged as a cornerstone of Intelligent Transportation Systems (ITS) for enabling autonomous driving and traffic safety~\cite{11328909, ref14},~\cite{11397193, 11278649, ref001, ref002, ref003, ref004, ref005, ref006, ref007, ref008, ref009, ref010, ref011, ref012, ref013, ref014, ref015, ref016, ref017, ref018, ref019, ref020}. While semantic communication offers a breakthrough beyond traditional bit-based paradigms by focusing on meaning extraction~\cite{ref16}, \cite{ref28}, it remains vulnerable in complex vehicular environments\cite{ref29, ref30}. Despite its inherent robustness at low Signal-to-Interference-plus-Noise Ratio (SINR)~\cite{ref27, 11409374}, high mobility and signal blockage in IoV scenarios induce deep fading and distortion, potentially causing semantic decoding failures\cite{11315125},~\cite{11023535, 10745538}. Such errors are intolerable for safety-critical instructions~\cite{wu2024cddm}, \cite{qin2023generalized}. Therefore, enhancing physical layer channel quality to maintain sufficient SINR is imperative for guaranteeing high-fidelity semantic understanding.

Reconfigurable Intelligent Surface (RIS) has emerged as a pivotal technology for enhancing wireless communication by reshaping the propagation environment via massive low-cost reflecting elements~\cite{ref17}, \cite{di2020smart}, \cite{huang2019reconfigurable}. However, conventional passive RIS suffers from severe multiplicative path loss~\cite{ref23}, rendering its gain negligible in long-distance IoV scenarios~\cite{ref24}. To overcome this bottleneck, Active RIS has been introduced~\cite{ref19}. By integrating reflection amplifiers, Active RIS amplifies the reflected signals to compensate for the propagation loss, thereby achieving superior energy efficiency and data rates~\cite{zhi2022active}, \cite{zhang2022active}. However, Active RIS is not flawless: on one hand, the active elements introduce additional thermal noise which is subsequently amplified~\cite{ref2}; on the other hand, the amplification gain is constrained by hardware costs and power budgets~\cite{lv2022ris}. This indicates that relying solely on power amplification leads to a fundamental performance bottleneck. Furthermore, conventional RIS elements are typically arranged in uniform arrays, whose geometric rigidity leads to rapidly saturating beamforming gains and upper-bounded Degrees of Freedom (DoF) as the array size scales~\cite{11143364, ref8}.

To breakthrough this spatial boundary, researchers have turned to exploiting the potential of the spatial domain. Inspired by fluid and movable antennas (MAs) \cite{ref20}, the Movable-Element RIS (ME-RIS) reconstructs the array geometry by adjusting element positions on rails, utilizing additional spatial diversity to break the geometric rigidity of traditional fixed arrays without increasing power consumption \cite{zhou2025star}, \cite{zhou2025movable}. Although shifting element positions can expand spatial DoFs to maximize achievable rates or enhance security~\cite{zhuang2025robust}, existing ME-RIS schemes rely almost exclusively on passive reflection.

To bridge these gaps, it is essential to analyze the individual limitations of active and movable RISs, and explain how vehicular semantic communications can uniquely benefit from their synergistic integration. First, while active RIS can mitigate multiplicative path loss, it introduces a fundamental ``noise amplification-power budget'' dilemma under practical hardware constraints~\cite{lv2022ris}. Crucially, its rigid, fixed array geometry lacks the agility to adaptively reconstruct the spatial propagation environment, making it unable to physically bypass localized deep fades in highly dynamic vehicular channels. Second, although movable RIS offers spatial diversity by shifting array geometries, existing state-of-the-art architectures rely almost exclusively on passive reflection~\cite{zhou2025movable}, \cite{zhuang2025robust}. In typical long-distance vehicle-to-infrastructure (V2I) environments, the severe multiplicative path loss entirely drowns out the spatial diversity gains offered by element mobility, failing to guarantee a sufficient SINR for reliable decoding.

Finally, semantic communication stands to uniquely benefit from this joint control due to its high vulnerability to physical-layer fading effects. To improve the transmission reliability of semantic features, some recent pioneering works have explored the integration of RIS or movable antennas with semantic communications. For instance, joint active and passive beamforming designs were proposed in~\cite{10614888} for RIS-aided semantic systems, and a cross-layer secure resource allocation scheme was developed in~\cite{10758370} for IRS-enhanced secure semantic networks. However, these frameworks rely on passive, fixed-position arrays, which can neither mitigate the severe cascaded path loss in long-distance V2I links nor physically realign elements to bypass localized deep fades. On the other hand, although fluid antennas have recently been introduced to assist near-field integrated sensing, computing, and semantic communication to exploit spatial diversity~\cite{11096081}, they focus primarily on transceiver-side mobility under passive propagation, leaving the cooperative active noise amplification and long-distance vehicular path loss unaddressed.

Indeed, because semantic systems transmit highly compressed features~\cite{ref3, ref7}, even a transient dip in SINR can trigger catastrophic decoding errors across an entire sentence~\cite{11315125}. Neither active amplification, which is bottlenecked by amplified thermal noise, nor element mobility, which suffers from severe path loss, can individually provide the stable, high-fidelity channel required for robust semantic parsing. In contrast, a joint active-mobility control framework establishes a cooperative physical-layer defense: active amplification elevates the average SINR to conquer long-distance path loss, while element mobility dynamically realigns the array to physically escape localized deep fades. When co-designed with adaptive semantic parameters~\cite{ref4}, this unified architecture provides a robust, flat, and fade-free physical-to-semantic transmission pipeline tailored to the demands of semantic-level understanding.

Motivated by these insights, this paper proposes a row-movable active RIS (RM-A-RIS) architecture and, more importantly, a unified co-design framework that jointly optimizes active amplification, element mobility, and semantic symbol adaptation for vehicular semantic communications. This joint optimization is characterized by a tightly coupled variable space. Specifically, when RIS elements become movable, their positions not only reshape the cascaded channel but also redistribute the amplified thermal noise in the spatial domain. The resulting SINR then interacts with the discrete semantic codebook through a highly nonlinear similarity function. Based on this, we rigorously model the non-trivial coupling of position-dependent active noise, spatial diversity bypass, and semantic spectral efficiency (SSE)\footnote{The source code is available at: \href{https://github.com/qiongwu86/Active-Movable-Element-RIS-Assisted-Vehicular-Semantic-Communications-Modeling-and-Optimization}{https://github.com/qiongwu86/Active-Movable-Element-RIS-Assisted-Vehicular-Semantic-Communications-Modeling-and-Optimization}.}. The main contributions are summarized as follows:
\begin{itemize}
	\item We propose a hardware-feasible RM-A-RIS-assisted vehicular semantic communication architecture and design a two-timescale chronological joint optimization protocol to reconcile continuous spatial optimization with physical mechanical latency. Under this protocol, slow-timescale mechanical adjustments of the RIS element positions are restricted to frame boundaries to adapt to the slowly changing vehicular topology, while low-complexity, fast-timescale adaptations for active phase shifts and semantic symbol lengths are executed slot-by-slot to track fast fading, thereby satisfying practical hardware constraints.
	
	\item To address the coupled non-convexity of the reflection coefficients under both total power and individual element saturation constraints, we develop a rigorous optimization pipeline. To resolve the non-differentiability of the discrete empirical semantic similarity lookup table, we construct a continuously differentiable surrogate function via cubic spline interpolation. Utilizing successive convex approximation (SCA) and fractional programming quadratic transform techniques, we reformulate the fractional SINR and derive an analytical, adaptive update rule for the active reflection coefficients, which strictly prevents amplifier saturation.
	
	\item For the continuous RIS element positions, we develop a Projected Gradient Ascent (PGA) algorithm combined with the Penalty Method to systematically solve the spatial deployment problem. To depart from standard heuristic search methods which suffer from exponential complexity and quantization errors, we transform the coupled ordering constraints and deploy the Pool Adjacent Violators Algorithm (PAVA) to achieve an exact Euclidean projection onto the ordered box constraints. Combined with a discrete greedy search for semantic symbol lengths, the proposed algorithm breaks the geometric rigidity of uniform arrays. Simulation results show substantial SSE improvements over state-of-the-art active, passive, and heuristic baselines.
\end{itemize}

The remainder of this paper is organized as follows. In Section~\ref{sec:system_model}, the system model of the RM-A-RIS assisted vehicular semantic communication is established, incorporating the position-dependent active noise, and the joint optimization problem for maximizing SSE is formulated. Section~\ref{sec:phase_opt} details the phase optimization algorithm, where the coupled non-convex objective is tackled via semantic sensitivity weighting and quadratic transform. Section~\ref{sec:position_opt} elaborates on the joint position and semantic symbol length optimization, utilizing the penalty-based PGA and greedy search strategies, respectively. Numerical results are presented in Section~\ref{sec:simulation} to demonstrate the superiority of the proposed scheme and validate the theoretical analysis. Finally, Section~\ref{sec:conclusion} concludes this paper.

\textit{Notations:} Scalars, vectors, and matrices are denoted by italic letters, boldface lower-case letters, and boldface upper-case letters, respectively. The operators $(\cdot)^T$, $(\cdot)^H$, and $|\cdot|$ denote the transpose, conjugate transpose, and absolute value (or modulus), respectively. $\|\cdot\|_2$ and $\|\cdot\|_F$ represent the Euclidean norm of a vector and the Frobenius norm of a matrix. $\mathbb{C}^{M \times N}$ and $\mathbb{R}^{M \times N}$ denote the spaces of $M \times N$ complex-valued and real-valued matrices, respectively. $\mathcal{CN}(\mathbf{0}, \mathbf{I})$ denotes the circularly symmetric complex Gaussian (CSCG) distribution with zero mean and identity covariance matrix $\mathbf{I}$. $\mathrm{diag}(\mathbf{x})$ represents a diagonal matrix with the elements of vector $\mathbf{x}$ on its main diagonal. $\mathrm{Re}\{\cdot\}$ denotes the real part of a complex number and $\mathbb{E}[\cdot]$ denotes the statistical expectation. $[\mathbf{Q}]_{ii}$ represent the $(i,i)$-th element in matrix $\mathbf{Q}$.
\section{System Model}
\label{sec:system_model}  
\subsection{Scene Geometry and Dynamic Position Modeling}
\begin{figure}[!t]
	\centering
	\includegraphics[width=3.5in]{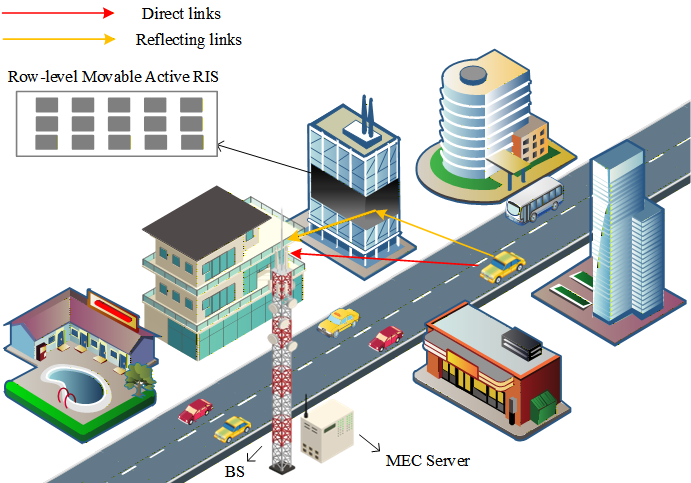}
	\caption{System model}
	\label{fig_1}
\end{figure}

We consider an uplink V2I communication scenario, as illustrated in Fig.~\ref{fig_1}. The system is established within a three-dimensional (3D) Cartesian coordinate system. The BS is equipped with a uniform linear array (ULA) consisting of $N_{BS}$ antennas and is located at a fixed coordinate $\mathbf{p}_{BS} = [x_{BS}, y_{BS}, z_{BS}]^T$ \cite{10847910, 11190066}. On the roadway, there are $K$ single-antenna vehicles distributed. The position of the $k$-th vehicle at time slot $t$ is denoted as $\mathbf{p}_k(t) = [x_k(t), y_k, 0]^T$.

To assist communication, a RM-A-RIS is deployed on the facade of a roadside building. 
The RIS comprises $M$ rows and $N$ columns, totaling $M \times N$ reflecting elements. 
To reduce hardware complexity, a row-level movable architecture is adopted, where all elements 
in the $m$-th row ($m \in \{1, \dots, M\}$) are mounted on the same horizontal guide rail of 
length $L_{rail}$ and share the same height $z_m$. Specifically, each of the $N$ reflecting elements on the same guide rail can slide and be positioned independently along the horizontal direction. The position vector $\mathbf{u}_{m,n} \in \mathbb{R}^3$ of the element 
in the $m$-th row and $n$-th column ($n \in \{1, \dots, N\}$) is defined as:
\begin{equation}
	\mathbf{u}_{m,n} = [x_{m,n}, y_{RIS}, z_m]^T,
\end{equation}
where $x_{m,n}$ represents the one-dimensional coordinate variable of the element along the guide rail. To ensure physical feasibility, the position variables must satisfy the guide rail boundary constraints and the collision avoidance constraints for adjacent elements:
\begin{equation}
	\mathcal{P} = \left\{ x_{m,n} \;\middle|\;
	\begin{aligned}
		& 0 \le x_{m,n} \le L_{rail}, \quad \forall m, n \\
		& x_{m,n+1} - x_{m,n} \ge D_{min}, \quad \forall m, 1 \le n < N
	\end{aligned}
	\right\},
\end{equation}
where $D_{min}$ denotes the minimum safety distance between adjacent elements. We denote the set of position coordinates of all elements as the matrix $\mathbf{U} \in \mathbb{R}^{M \times N}$, where $[\mathbf{U}]_{m,n} = x_{m,n}$.
\subsection{Channel Model}
Since the RIS element positions $\mathbf{U}$ are continuously adjustable, the channel state information (CSI) is an explicit function of $\mathbf{U}$. We adopt a geometric channel model to capture the impact of position variations on large-scale fading and phase shifts. To realistically reflect the complex urban vehicular environment characterized by frequent blockages and scattering, we model the vehicle-to-BS direct links as Rayleigh fading, the vehicle-to-RIS links as Rician fading, and the elevated static RIS-to-BS links as Line-of-Sight (LoS) dominated channels. The specific channel matrices are modeled as follows.

\subsubsection{Vehicle-to-RIS Channel}
Due to the presence of both direct line-of-sight and local scatterers around the vehicles, the vehicle-to-RIS link is modeled as a Rician fading channel. The Euclidean distance between the $k$-th vehicle and the $(m, n)$-th element of the RIS is given by:
\begin{equation}
	\begin{split}
		d_{sr,k}^{m,n}(\mathbf{U}) &= \|\mathbf{u}_{m,n} - \mathbf{p}_k\|_2 \\
		&= \sqrt{(x_{m,n} - x_k)^2 + (y_{RIS} - y_k)^2 + z_m^2}.
	\end{split}
\end{equation}
The corresponding channel coefficient $h_{sr, k}^{m,n}(\mathbf{U})$ is determined by the distance-dependent path loss, log-normal shadowing, and the Rician multipath components:
\begin{equation}
	\begin{split}
		h_{sr, k}^{m,n}(\mathbf{U}) &= \sqrt{\beta_0 (d_{sr, k}^{m,n}(\mathbf{U}))^{-\alpha_{sr}} \zeta_{sr,k}} \\
		&\quad \times \left( \sqrt{\frac{K_{sr}}{K_{sr}+1}} \bar{h}_{sr,k}^{m,n}(\mathbf{U}) + \sqrt{\frac{1}{K_{sr}+1}} \tilde{h}_{sr,k} \right),
	\end{split}
\end{equation}
where $\beta_0$ denotes the path loss at the reference distance, $\alpha_{sr}$ is the path loss exponent, $\zeta_{sr,k}$ is the shadow fading coefficient with $10 \log_{10} \zeta_{sr,k} \sim \mathcal{N}(0, \sigma_{sh}^2)$, and $K_{sr}$ is the Rician $K$-factor. The deterministic LoS phase shift is given by $\bar{h}_{sr,k}^{m,n}(\mathbf{U}) = e^{-j \frac{2\pi}{\lambda} d_{sr, k}^{m,n}(\mathbf{U})}$, and $\lambda$ represents the carrier wavelength. The variable $\tilde{h}_{sr,k} \sim \mathcal{CN}(0,1)$ represents the normalized NLoS small-scale scattering component. The complete vehicle-to-RIS channel vector $\mathbf{h}_{sr, k}(\mathbf{U}) \in \mathbb{C}^{MN \times 1}$ is defined as the column stack of the channel coefficients of all elements:
\begin{equation}
	\begin{split}
		\mathbf{h}_{sr,k}(\mathbf{U}) \stackrel{\Delta}{=} \Big[ & h_{sr,k}^{(11)}(\mathbf{U}), \dots, h_{sr,k}^{(1N)}(\mathbf{U}), \dots, \\
		& h_{sr,k}^{(M1)}(\mathbf{U}), \dots, h_{sr,k}^{(MN)}(\mathbf{U}) \Big]^T.
	\end{split}
\end{equation}

\subsubsection{RIS-to-BS Channel}
Since both the RM-A-RIS and the BS antennas are typically deployed at elevated positions (e.g., on building facades and masts) with clear clearances, we assume the RIS-to-BS link is free from random scattering and blockages, and is thus dominated by the deterministic LoS path. The distance from the $(m, n)$-th element of the RIS to the center of the BS is given by $d_{rd}^{(m,n)}(\mathbf{U}) = \|\mathbf{p}_{BS} - \mathbf{u}_{m,n}\|_2$. Let $\vartheta_{m,n}$ denote the cosine of the Angle of Arrival (AoA) at the BS corresponding to this element. The array response vector at the BS, $\mathbf{a}_{BS}(\vartheta_{m,n}) \in \mathbb{C}^{N_{BS} \times 1}$, is defined as:
\begin{equation}
	\mathbf{a}_{BS}(\vartheta_{m,n}) = \left[ 1, e^{j\pi\vartheta_{m,n}}, \dots, e^{j(N_{BS}-1)\pi\vartheta_{m,n}} \right]^T.
\end{equation}
Then, the sub-channel vector from the $(m, n)$-th element of the RIS to the BS, $\mathbf{h}_{rd}^{(m,n)}(\mathbf{U}) \in \mathbb{C}^{N_{BS} \times 1}$, is expressed as:
\begin{equation}
	\begin{split}
		&\mathbf{h}_{rd}^{(m,n)}(\mathbf{U}) = \\& \sqrt{\beta_0 \left(d_{rd}^{(m,n)}(\mathbf{U})\right)^{-\alpha_{rd}}} e^{-j\frac{2\pi}{\lambda} d_{rd}^{(m,n)}(\mathbf{U})} \mathbf{a}_{BS}(\vartheta_{m,n}),
	\end{split}
\end{equation}
where $\alpha_{rd}$ denotes the path loss exponent of the RIS-to-BS link. The complete RIS-BS channel matrix $\mathbf{H}_{rd}(\mathbf{U}) \in \mathbb{C}^{N_{BS} \times MN}$ is defined as the column concatenation of all sub-channel vectors:
\begin{equation}
	\begin{split}
		\mathbf{H}_{rd}(\mathbf{U}) \stackrel{\Delta}{=} \bigg[ & \mathbf{h}_{rd}^{(1,1)}(\mathbf{U}), \dots, \mathbf{h}_{rd}^{(1,N)}(\mathbf{U}), \dots, \\
		& \mathbf{h}_{rd}^{(M,1)}(\mathbf{U}), \dots, \mathbf{h}_{rd}^{(M,N)}(\mathbf{U}) \bigg].
	\end{split}
\end{equation}

\subsubsection{Direct Link}
The direct channel between the $k$-th vehicle and the BS, denoted as $\mathbf{h}_{d,k} \in \mathbb{C}^{N_{BS} \times 1}$, is modeled as a composite channel incorporating large-scale path loss, log-normal shadow fading, and NLoS Rayleigh small-scale fading. It is expressed as:
\begin{equation}
	\mathbf{h}_{d,k} = \sqrt{\beta_0 \|\mathbf{p}_{BS} - \mathbf{p}_k\|^{-\alpha_d} \zeta_{d,k}} \tilde{h}_{d,k} \mathbf{a}_{BS}(\theta_k),
\end{equation}
where $\alpha_d$ is the path loss exponent of the direct link, and $\zeta_{d,k}$ represents the shadow fading coefficient which follows a log-normal distribution, i.e., $10 \log_{10} \zeta_{d,k} \sim \mathcal{N}(0, \sigma_{sh}^2)$. The variable $\tilde{h}_{d,k} \sim \mathcal{CN}(0, 1)$ denotes the normalized small-scale fading coefficient, and $\mathbf{a}_{BS}(\theta_k)$ is the steering vector at the BS corresponding to the AoA $\theta_k$ of the direct link from the $k$-th vehicle.
\subsection{Active RIS Signal Transmission and Noise Model}

Unlike passive RIS, active RIS amplifies the reflected signals while inevitably introducing thermal noise \cite{ref2}.

\subsubsection{Signal Amplification at the RIS}

Let $\mathbf{v} = [v_1, \dots, v_{MN}]^T \in \mathbb{C}^{MN \times 1}$ denote the active reflection coefficient vector, and the corresponding diagonal amplification matrix be $\mathbf{\Psi} = \mathrm{diag}(\mathbf{v})$. The incident signal arriving at the RIS is superimposed with the equivalent baseband thermal noise $\mathbf{n}_R \sim \mathcal{CN}(\mathbf{0}, \sigma_R^2 \mathbf{I}_{MN})$ introduced by the RIS, where the noise power per element is physically modeled as $\sigma_R^2 = k_B T B F_{\text{RIS}}$. Here, $k_B$ denotes the Boltzmann constant, $T$ is the equivalent noise temperature, $B$ represents the system bandwidth, and $F_{\text{RIS}}$ is the noise figure of the active reflection amplifiers. The signal reflected and amplified by the RIS, $\mathbf{x}_R \in \mathbb{C}^{MN \times 1}$, is expressed as:
\begin{equation}
	\mathbf{x}_R = \mathbf{\Psi} \left( \sum_{k=1}^{K} \mathbf{h}_{sr,k}(\mathbf{U}) \sqrt{P_k} s_k + \mathbf{n}_R \right),
\end{equation}
where $P_k$ denotes the transmit power of the $k$-th vehicle, and $s_k \sim \mathcal{CN}(0, 1)$ represents the normalized transmitted symbol containing semantic information, satisfying $\mathbb{E}[|s_k|^2] = 1$.
\subsubsection{BS Received Signal and SINR}

The signal received at BS is the superposition of direct signal and signal forwarded by the RIS. To simplify the expression, we define the total effective channel vector from the $k$-th vehicle to BS as:
\begin{equation}
	\mathbf{h}_{eff, k}(\mathbf{U}, \mathbf{v}) \stackrel{\Delta}{=} \mathbf{h}_{d,k} + \mathbf{H}_{rd}(\mathbf{U}) \mathrm{diag}(\mathbf{h}_{sr, k}(\mathbf{U}))\mathbf{v}.
\end{equation}
Note that $\mathrm{diag}(\mathbf{h}_{sr, k}(\mathbf{U}))\mathbf{v}$ is equivalent to $\mathbf{\Psi}\mathbf{h}_{sr, k}(\mathbf{U})$.

We assume that the system adopts Frequency Division Multiple Access (FDMA), where different vehicles occupy orthogonal Resource Blocks (RBs). For the $k$-th vehicle, the received signal $\mathbf{y}_k \in \mathbb{C}^{N_{BS} \times 1}$ at the BS contains only the useful signal of that vehicle, the active noise introduced and amplified by the RIS, and the thermal noise at the BS receiver:
\begin{equation}
	\begin{split}
		\mathbf{y}_k = & \underbrace{\sqrt{G_{veh}G_{bs}} \left( \mathbf{h}_{d,k} + \mathbf{H}_{rd}(\mathbf{U})\mathbf{\Psi}\mathbf{h}_{sr,k}(\mathbf{U}) \right) \sqrt{P_k} s_k}_{\text{Received useful signal}} \\
		& + \underbrace{\sqrt{G_{bs}} \mathbf{H}_{rd}(\mathbf{U})\mathbf{\Psi}\mathbf{n}_R}_{\text{Received active noise}} + \mathbf{n}_{BS},
	\end{split}
\end{equation}
where $G_{veh}$ and $G_{bs}$ denote the antenna gains of the vehicle and the BS, respectively. The variable $\mathbf{n}_{BS} \sim \mathcal{CN}(\mathbf{0}, \sigma_{total}^2 \mathbf{I}_{N_{BS}})$ represents the total thermal noise at the BS. Due to the orthogonality in the frequency domain, signals from other vehicles are filtered out and do not constitute interference. The BS employs a Maximum Ratio Combining (MRC) receiver to maximize the received SINR. Let $\mathbf{h}_{eff,k}$ denote the aggregate effective channel vector from vehicle $k$ to the BS, encompassing both the direct path and the RIS-reflected path. The beamforming vector for vehicle $k$ is given by:
\begin{equation} \label{eq:wk_mrc}
	\mathbf{w}_k = \frac{\mathbf{h}_{eff,k}}{\|\mathbf{h}_{eff,k}\|_2}.
\end{equation}
At each AO iteration, $\mathbf{w}_k$ is dynamically updated based on the current realization of the effective channel $\mathbf{h}_{eff,k}$, which depends on the updated RIS element positions $\mathbf{U}$ and active reflection coefficients $\mathbf{v}$. Consequently, received SINR after combining is expressed as:
\begin{equation}
	\gamma_k(\mathbf{U}, \mathbf{v}) = \frac{G_{veh} G_{bs} P_k \left| \mathbf{w}_k^H \mathbf{h}_{eff,k}(\mathbf{U}, \mathbf{v}) \right|^2}{G_{bs} \sigma_R^2 \left\| \mathbf{w}_k^H \mathbf{H}_{rd}(\mathbf{U}) \mathbf{\Psi} \right\|_2^2 + \sigma_{total}^2}.
\end{equation}
The first term in the denominator, $\sigma_R^2 \|\mathbf{w}_k^H \mathbf{H}_{rd}(\mathbf{U})\mathbf{\Psi}\|_2^2$ reveals how the active noise power is jointly influenced by the RIS element positions $\mathbf{U}$ and amplification gain matrix $\mathbf{\Psi}$.
\subsubsection{Active RIS Power Consumption Model}

The total power consumption of the active RIS is constrained by the capability of the hardware amplifiers. The total output power of the RIS, denoted as $P_{total}$, is defined as the sum of the reflected signal power and the amplified noise power, calculated as follows:
\begin{equation}
	P_{total}(\mathbf{U}, \mathbf{v}) = \sum_{k=1}^{K} P_k \| \mathbf{\Psi} \mathbf{h}_{sr,k}(\mathbf{U}) \|_2^2 + \sigma_R^2 \| \mathbf{v} \|_2^2.
\end{equation}
This output power must satisfy the constraint $P_{total} \le P_{max}^{RIS}$, where $P_{max}^{RIS}$ represents the maximum allowable output power of the active RIS.

\subsection{Semantic Transmission and Spectral Efficiency}
We adopt a feature extraction-based semantic communication model \cite{ref3}. Assume that the semantic task transmitted by the $k$-th vehicle consists of $L$ words, with each word containing an average semantic information amount of $I$ (suts). The semantic encoder compresses each word into $q_k$ physical layer symbols. Consequently, the semantic transmission rate $\Phi_k$ (sut/s) is defined as \cite{ref4}:
\begin{equation}
	\Phi_k = \frac{B \cdot I}{q_k L} \xi_k(\gamma_k, q_k),
\end{equation}
where $B$ denotes the bandwidth, and $\xi_k(\gamma_k, q_k) \in [0,1]$ represents the semantic similarity function. Note that $\xi_k$ is a complex nonlinear function of the received SINR $\gamma_k$ and the encoding length $q_k$.

To explicitly define and obtain $\xi_k$, we follow the Deep Learning enabled Semantic Communication (DeepSC) framework \cite{ref4}, where the underlying semantics of the text are extracted and recovered using Transformer-based semantic encoders and decoders. To accurately measure the semantic fidelity, a pre-trained Sentence-Bidirectional Encoder Representations from Transformers (Sentence-BERT) model is utilized to evaluate the BERT-level semantic similarity between the transmitted sentence $s$ and the decoded sentence $\hat{s}$. Specifically, the similarity metric is calculated as the cosine similarity of their sentence embedding vectors, given by $\xi = \frac{\mathbf{B}(s)\mathbf{B}(\hat{s})^T}{\|\mathbf{B}(s)\| \|\mathbf{B}(\hat{s})\|}$, where $\mathbf{B}(\cdot)$ denotes the Sentence-BERT embedding function. 

Since analytical expressions for $\xi_k$ are mathematically intractable, it is practically obtained via a lookup table. To generate this table, the well-trained DeepSC model is extensively evaluated over an Additive White Gaussian Noise (AWGN) channel using a specific text corpus. By simulating the transmission under various discrete combinations of SINR $\gamma_k$ and encoding length $q_k$, the statistical averages of the resulting BERT-level similarities are recorded. For continuous channel conditions in practical resource allocation, the value of $\xi_k(\gamma_k, q_k)$ is then acquired using two-dimensional cubic spline interpolation (also known as bicubic spline interpolation) based on this pre-generated lookup table. This interpolation method not only ensures unbiased exact matches at the discrete empirical grid points, but also constructs a globally smooth and twice-differentiable surface, which provides a rigorous mathematical foundation for the gradient-based optimization in the subsequent sections.

Based on the above semantic transmission model, we define the semantic spectral efficiency (SSE) as:
\begin{equation} \label{eq:sse_def}
	 SSE_k(\gamma_k, q_k) \triangleq \frac{\Phi_k}{B} = \frac{I}{q_k L} \xi_k(\gamma_k, q_k) \quad (\mathrm{suts/s/Hz}).
\end{equation}
\vspace{-0.95cm}
\subsection{Channel Aging Due to Mobility}
In practical high-mobility scenarios, the BS cannot acquire perfect instantaneous CSI. Considering the Doppler shift $f_{d,k} = v_k/\lambda$ induced by the vehicle velocity $v_k$, as well as the system processing delay $\tau$, the channel suffers from the aging phenomenon \cite{11119330}. According to Jakes' correlation model, the temporal correlation coefficient for the $k$-th vehicle is given by $\rho_{k,X} = J_0(2\pi f_{d,k} \cos(\theta_{k,X}) \tau)$, where $\theta_{k,X}$ denotes the angle between the vehicle's velocity vector and the propagation direction towards node $X \in \{\text{BS, RIS}\}$. 
Based on our channel modeling, the estimated small-scale fading components available at the BS for optimization are defined as:
	\begin{align}
		\hat{h}_{d,k} &= \rho_{k,BS} \tilde{h}_{d,k} + \sqrt{1 - \rho_{k,BS}^2} e_d, \\
		\hat{h}_{sr,k} &= \rho_{k,RIS} \tilde{h}_{sr,k} + \sqrt{1 - \rho_{k,RIS}^2} e_{sr},
	\end{align}
where $e_d, e_{sr} \sim \mathcal{CN}(0,1)$ represent the independent Gaussian estimation errors. 
It is pivotal to note that the CSI aging solely degrades the NLoS scattering components ($\tilde{h}_{d,k}$ and $\tilde{h}_{sr,k}$). For the vehicle-to-RIS Rician link, the geometric LoS component $\bar{h}_{sr,k}^{m,n}(\mathbf{U})$ is determined instantaneously by the continuous position mapping, thereby evading the aging effect. Furthermore, since the elevated RIS-to-BS link $\mathbf{H}_{rd}(\mathbf{U})$ is purely geometric and devoid of random scatterers, it is deterministic and perfectly known once the positions $\mathbf{U}$ are given. 
In the proposed framework, the BS performs the optimization of RIS element positions $\mathbf{U}$ and phase shifts $\mathbf{v}$ based on the outdated effective channel constructed from $\hat{h}_{d,k}$ and $\hat{h}_{sr,k}$. However, the actual system performance is realistically evaluated based on the true instantaneous channel state.
\subsection{Optimization Problem Formulation}

Our goal is to jointly optimize the element positions $\mathbf{U}$ of the RM-A-RIS, the active reflection coefficients $\mathbf{v}$, and the semantic symbol lengths $\mathbf{q} = [q_1, \dots, q_K]^T$, to maximize the Sum-SSE. The optimization problem is formulated as:

\begin{subequations} \label{prob:original}
	\begin{align}
		\max_{\mathbf{U}, \mathbf{v}, \mathbf{q}} \quad & \sum_{k=1}^{K} \frac{I}{q_k L} \xi_k(\gamma_k(\mathbf{U}, \mathbf{v}), q_k) \label{obj:sum_sse} \\
		\textrm{s.t.} \quad & \xi_k \ge \xi_{th}, \quad \forall k \in \mathcal{K}, \label{const:qos} \\
		& \sum_{k=1}^{K} P_k \|\mathrm{diag}(\mathbf{h}_{sr,k}(\mathbf{U}))\mathbf{v}\|_2^2 + \sigma_R^2 \|\mathbf{v}\|_2^2 \le P_{max}^{RIS}, \label{const:power} \\
		& |v_i|^2 \left( \sum_{k=1}^{K} P_k |[\mathbf{h}_{sr,k}(\mathbf{U})]_i|^2 + \sigma_R^2 \right) \le P_{sat},  \label{const:sat} \\
		& |v_i| \le A_{max}, \quad \forall i \in \{1, \dots, MN\}, \label{const:amp} \\
		& q_k \in \{1, \dots, Q_{max}\}, \quad \forall k \in \mathcal{K}, \label{const:discrete} \\
		& \mathbf{U} \in \mathcal{P}. \label{const:pos}
	\end{align}
\end{subequations}

Constraint \eqref{const:qos} ensures the minimum semantic similarity requirement for the vehicles; Constraint \eqref{const:power} limits the total output power of the active RIS; Constraint \eqref{const:sat} restricts the output power of each individual reflection amplifier to prevent non-linear saturation, where $P_{sat}$ denotes the saturation output power of a single active element; Constraint \eqref{const:amp} restricts the maximum amplification gain amplitude of amplifiers; Constraint \eqref{const:discrete} confines the semantic symbol length to discrete integers; Constraint \eqref{const:pos} defines the geometric feasible region of the RM-A-RIS. This is a highly non-convex Mixed-Integer Non-Linear Programming (MINLP) problem. We employ an AO algorithm based on fractional programming and projected gradient methods to solve it.

\section{Active Phase Optimization}
\label{sec:phase_opt}     
To address this challenging non-convex mixed-integer programming problem, we propose an AO framework to obtain a suboptimal solution with low computational complexity. First, we fix the positions of RIS elements and set of semantic symbol lengths $\mathbf{q}$. Consequently, the original problem degenerates into a sub-problem optimizing the active RIS reflection coefficient vector $\mathbf{v} = \mathrm{diag}(\mathbf{\Psi})$. Our objective is to maximize the system SSE (Sum-SSE) subject to the maximum power and amplitude constraints.
\subsection{SCA-based Semantic Sensitivity Approximation}

Due to the complex non-linear coupling between the semantic similarity function $\xi_k(\gamma_k)$ and the SINR $\gamma_k$, direct optimization is intractable. Moreover, since $\xi_k(\gamma_k)$ is inherently obtained from discrete lookup tables, we apply cubic spline interpolation to fit these empirical data points into a continuously differentiable function. This smooth approximation not only ensures the global continuity of the first and second-order derivatives but also enables the exact computation of analytical gradients, thereby guaranteeing the convergence and stability of the SCA algorithm. Based on the SCA theory, we perform a first-order Taylor expansion of the semantic spectral efficiency function with respect to $\gamma_k$ at the $t$-th iteration. We construct the following surrogate optimization objective:
\begin{equation}
	\max_{\mathbf{v}} \sum_{k=1}^{K} \alpha_k^{(t)} \gamma_k(\mathbf{v}),
\end{equation}
where $\alpha_k^{(t)}$ is defined as the semantic sensitivity weight for vehicle $k$, representing the analytical gradient of the semantic efficiency with respect to the physical layer SINR:
\begin{equation}
	 \alpha_k^{(t)} = \frac{\partial SSE_k}{\partial \gamma_k} \bigg|_{\gamma_k=\gamma_k^{(t)}} = \frac{I}{q_k L} \frac{\partial \xi_k(\gamma_k)}{\partial \gamma_k} \bigg|_{\gamma_k=\gamma_k^{(t)}},
\end{equation}
where the partial derivative $\partial \xi_k / \partial \gamma_k$ is precisely calculated using the analytical derivative of the fitted cubic spline polynomials. Furthermore, the physical layer SINR is given by:
\begin{equation} \label{23}
	\gamma_k(\mathbf{v}) = \frac{G_{veh} G_{bs} P_k \left| \mathbf{w}_k^H \mathbf{h}_{eff, k}(\mathbf{v}) \right|^2}{G_{bs} \sigma_R^2 \left\| \mathbf{w}_k^H \mathbf{H}_{rd} \mathbf{\Psi} \right\|_2^2 + \sigma_{total}^2},
\end{equation}
where the vector $\mathbf{v} = \mathrm{diag}(\mathbf{\Psi}) \in \mathbb{C}^{MN \times 1}$ denotes the active RIS reflection coefficients. Specifically, the receive beamforming vectors $\{\mathbf{w}_k\}$ in \eqref{23} are assumed to be fixed to their values from the previous iteration, and are updated only after the active reflection coefficient vector $\mathbf{v}$ is optimized. Given fixed RIS positions $\mathbf{U}$, we reformulate the numerator (signal power) and denominator (active noise power) of the original SINR expression into vector forms.
\subsubsection{Vectorization of Useful Signal Power}

Based on the property of diagonal matrix multiplication $\mathbf{A}\mathrm{diag}(\mathbf{b})\mathbf{c} = \mathbf{A}\mathrm{diag}(\mathbf{c})\mathbf{b}$, we rewrite the cascaded channel as a linear form with respect to $\mathbf{v}$, explicitly preserving the direct link as a constant offset. The amplitude of the received signal for vehicle $k$ is derived as follows:
\begin{equation}
	\begin{split}
		|\tilde{y}_{sig,k}| & = \sqrt{P_k G_{sys}} \left| \mathbf{w}_k^H (\mathbf{h}_{d,k} + \mathbf{H}_{rd}\mathbf{\Psi}\mathbf{h}_{sr,k}) \right| \\
		& = \sqrt{P_k G_{sys}} \left| \mathbf{w}_k^H \mathbf{h}_{d,k} + \mathbf{w}_k^H \mathbf{H}_{rd} \mathrm{diag}(\mathbf{h}_{sr,k})\mathbf{v} \right| \\
		& = \Bigg| \underbrace{\sqrt{P_k G_{sys}}\mathbf{w}_k^H \mathbf{h}_{d,k}}_{b_k} \\[-10pt] 
		& \quad + \underbrace{\sqrt{P_k G_{sys}}\left(\mathrm{diag}(\mathbf{h}_{sr,k})\mathbf{H}_{rd}^H\mathbf{w}_k\right)^H}_{\mathbf{a}_k^H} \mathbf{v} \Bigg| \\[-3pt] 
		& = |\mathbf{a}_k^H \mathbf{v} + b_k|,
	\end{split}
\end{equation}
where $G_{sys} = G_{veh}G_{bs}$ denotes the total system antenna gain. Here, $b_k \in \mathbb{C}$ represents the effective direct channel gain, which is a constant scalar, and $\mathbf{a}_k \in \mathbb{C}^{MN \times 1}$ represents the effective cascaded channel vector. Consequently, the SINR numerator is reformulated as an affine squared modulus form $|\mathbf{a}_k^H \mathbf{v} + b_k|^2$ with respect to $\mathbf{v}$.
\subsubsection{Quadratic Formulation of Active Noise Power}

The active noise term in the denominator originates from the thermal noise introduced by RIS, which is subsequently processed by beamforming vector. Utilizing the identity $\|\mathbf{A}\mathrm{diag}(\mathbf{x})\|_F^2 = \mathbf{x}^H \mathrm{diag}(\mathbf{A}^H \mathbf{A}) \mathbf{x}$, the derivation is as follows:
\begin{equation}
	\begin{split}
		P_{active,k} & = G_{bs}\sigma_R^2 \left\| \mathbf{w}_k^H \mathbf{H}_{rd} \mathbf{\Psi} \right\|_2^2 \\
		& = G_{bs}\sigma_R^2 \sum_{i=1}^{MN} \left| [\mathbf{w}_k^H \mathbf{H}_{rd}]_i \cdot v_i \right|^2 \\
		& = \mathbf{v}^H \underbrace{\left( G_{bs}\sigma_R^2 \cdot \mathrm{diag}\left( \left| \mathbf{H}_{rd}^H \mathbf{w}_k \right|^2 \right) \right)}_{\mathbf{Q}_k} \mathbf{v},
	\end{split}
\end{equation}
where $\mathbf{Q}_k \in \mathbb{R}^{MN \times MN}$ is defined as the active noise covariance matrix for vehicle $k$. As indicated by the derivation, $\mathbf{Q}_k$ is a diagonal matrix.

Combining the above derivations, the SINR for vehicle $k$ can be reformulated into a standard fractional form with respect to $\mathbf{v}$:
\begin{equation}
	\gamma_k(\mathbf{v}) = \frac{|\mathbf{a}_k^H \mathbf{v} + b_k|^2}{\mathbf{v}^H \mathbf{Q}_k \mathbf{v} + \sigma_{total}^2}.
\end{equation}
\subsection{Constraint Analysis}
Next, we analyze the constraints. The semantic similarity function in constraint \eqref{const:qos} is a mapping based on measurement data and lacks a specific analytical expression; thus, its convexity cannot be directly evaluated. Since the semantic similarity is jointly influenced by the SINR and $q_k$, and generally, a higher SINR and larger $q_k$ yield higher semantic similarity, we temporarily relax this constraint during the phase optimization stage. Instead, we ensure this constraint is satisfied during the subsequent optimization of the $q_k$ values.

Constraint \eqref{const:power} represents a convex constraint. First, the function is reformulated into a standard quadratic form. Let $\mathbf{D}_k = \mathrm{diag}(\mathbf{h}_{sr,k})$. Since $\mathbf{h}_{sr,k}$ is a complex vector, $\mathbf{D}_k$ is a complex diagonal matrix. Thus,
\begin{subequations}
	\begin{align}
		\| \mathrm{diag}(\mathbf{h}_{sr,k})\mathbf{v} \|_2^2 &= \| \mathbf{D}_k \mathbf{v} \|_2^2 \nonumber \\
		&= (\mathbf{D}_k \mathbf{v})^H (\mathbf{D}_k \mathbf{v}) \nonumber \\
		&= \mathbf{v}^H \mathbf{D}_k^H \mathbf{D}_k \mathbf{v}, \\
		\| \mathbf{v} \|_2^2 &= \mathbf{v}^H \mathbf{v} = \mathbf{v}^H \mathbf{I}_{MN} \mathbf{v},
	\end{align}
\end{subequations}
where $\mathbf{I}_{MN}$ denotes the identity matrix of size $MN \times MN$.

By aggregating these terms, the total power function $f(\mathbf{v})$ is derived as:
\begin{equation}
	\begin{split}
		f(\mathbf{v}) &= \sum_{k=1}^{K} P_k (\mathbf{v}^H \mathbf{D}_k^H \mathbf{D}_k \mathbf{v}) + \sigma_R^2 (\mathbf{v}^H \mathbf{I}_{MN} \mathbf{v}) \\
		&= \mathbf{v}^H \underbrace{\left( \sum_{k=1}^{K} P_k \mathbf{D}_k^H \mathbf{D}_k + \sigma_R^2 \mathbf{I}_{MN} \right)}_{\mathbf{Q}} \mathbf{v}.
	\end{split}
\end{equation}
Consequently, the constraint function is simplified into a standard quadratic form:
\begin{equation}
	f(\mathbf{v}) = \mathbf{v}^H \mathbf{Q} \mathbf{v}.
\end{equation}

According to convex optimization theory, the quadratic function $f(\mathbf{v})$ is convex if the matrix $\mathbf{Q}$ is Positive Semi-Definite (PSD). Since $\mathbf{D}_k = \mathrm{diag}(\mathbf{h}_{sr,k})$, the product $\mathbf{D}_k^H \mathbf{D}_k$ results in a real diagonal matrix where the diagonal elements correspond to the squared magnitudes of the channel gains, i.e., $[\mathbf{D}_k^H \mathbf{D}_k]_{ii} = |[\mathbf{h}_{sr,k}]_i|^2 \ge 0$. Therefore, $\mathbf{D}_k^H \mathbf{D}_k$ is PSD.

Given that the vehicle transmit power $P_k > 0$ and the noise power $\sigma_R^2 > 0$ are positive real numbers, and the identity matrix $\mathbf{I}_{MN}$ is Positive Definite (PD), the term $\sigma_R^2 \mathbf{I}_{MN}$ is strictly PD. The matrix $\mathbf{Q}$ is constructed as the sum of multiple PSD matrices and one PD matrix:
\begin{equation}
	\mathbf{Q} = \underbrace{\sum_{k=1}^{K} P_k \mathbf{D}_k^H \mathbf{D}_k}_{\text{PSD}} + \underbrace{\sigma_R^2 \mathbf{I}_{MN}}_{\text{PD}}.
\end{equation}

The sum of a Positive Semi-Definite matrix and a Positive Definite matrix is strictly PD. Therefore, $\mathbf{Q}$ is a PD matrix, which implies that constraint \eqref{const:power} is a convex constraint with respect to the variable $\mathbf{v}$.

Constraint \eqref{const:amp} is an inequality constraint, which is evidently convex.

Consequently, problem \eqref{obj:sum_sse} is transformed into a weighted sum-SINR maximization problem, formulated as follows:
\begin{subequations} \label{prob:optimization_final}
	\begin{align}
		\max_{\mathbf{v}} \quad & \sum_{k=1}^{K} \alpha_k^{(t)} \frac{|\mathbf{a}_k^H \mathbf{v} + b_k|^2}{\mathbf{v}^H \mathbf{Q}_k \mathbf{v} + \sigma_{total}^2} \label{eq:obj_30a} \\
		\textrm{s.t.} \quad & \mathbf{v}^H \mathbf{Q} \mathbf{v} \le P_{max}^{RIS}, \label{eq:const_30b} \\
		& |v_i| \le A_{max, i}^{eff}, \quad \forall i \in \{1, \dots, MN\}. \label{eq:const_30c}
	\end{align}
\end{subequations}

where $A_{max, i}^{eff}$ is the adaptive effective maximum gain amplitude for the $i$-th active element, which is forward-referenced and formally derived in \eqref{eq_eff_amp} to simultaneously satisfy both \eqref{const:amp} and \eqref{const:sat} under fixed $\mathbf{U}$. The aforementioned objective function $f(\mathbf{v})$ represents a weighted sum of multiple fractional quadratic functions, classifying the problem as a typical non-convex NP-hard problem. To address this, the Quadratic Transform method from fractional programming theory is employed. According to this theory, each fractional term in the form of $\frac{|A(\mathbf{v})|^2}{B(\mathbf{v})}$ can be reformulated by introducing an auxiliary variable $y$ into the form of $2\mathrm{Re}\{y^* A(\mathbf{v})\} - |y|^2 B(\mathbf{v})$.

An auxiliary variable $y_k \in \mathbb{C}$ is introduced for each vehicle $k$ to reformulate the original objective function \eqref{eq:obj_30a} into a new surrogate objective function with respect to $(\mathbf{v}, \mathbf{y})$:

\begin{equation} \label{eq:new_obj_g_vy}
	\begin{split}
		g(\mathbf{v}, \mathbf{y}) = \sum_{k=1}^{K} \Bigg( & 2\mathrm{Re} \left\{ y_k^* \sqrt{\alpha_k^{(t)}} (\mathbf{a}_k^H \mathbf{v} + b_k) \right\} \\
		& - |y_k|^2 \left( \mathbf{v}^H \mathbf{Q}_k \mathbf{v} + \sigma_{total}^2 \right) \Bigg).
	\end{split}
\end{equation}

This function exhibits the property of alternating convexity, allowing for the iterative optimization of $\mathbf{y}$ (with fixed $\mathbf{v}$) and $\mathbf{v}$ (with fixed $\mathbf{y}$).

First, we fix $\mathbf{v}$ to solve for the optimal $y_k$. The optimization term for each vehicle can be written as:
\begin{equation}\label{eq32}
	f(y_k) = 2\mathrm{Re}\{ y_k^* \sqrt{\alpha_k^{(t)}} A \} - |y_k|^2 B,
\end{equation}
where $A = \mathbf{a}_k^H \mathbf{v} + b_k$, and $B = \mathbf{v}^H \mathbf{Q}_k \mathbf{v} + \sigma_{total}^2$, both of which are constants at this step.

According to Wirtinger Calculus, to find the extremum, we take the partial derivative with respect to the conjugate of $y$, denoted as $y^*$:

\begin{equation}
	\frac{\partial f}{\partial y^*} = \sqrt{\alpha_k^{(t)}}A - yB.
\end{equation}

Setting the derivative to 0:
\begin{equation}
	\sqrt{\alpha_k^{(t)}}A - yB = 0,
\end{equation}
\begin{equation} \label{eq35}
	y^\star = \frac{\sqrt{\alpha_k^{(t)}}A}{B} = \frac{\sqrt{\alpha_k^{(t)}}(\mathbf{a}_k^H \mathbf{v} + b_k)}{\mathbf{v}^H \mathbf{Q}_k \mathbf{v} + \sigma_{total}^2}. \quad 
\end{equation}

After obtaining $y^\star$, we fix $y^\star$ to solve for the optimal value of $\mathbf{v}$. By substituting $y^\star$ back into the surrogate function, the first term can be expanded as $2\mathrm{Re}\{ y_k^\star \sqrt{\alpha_k^{(t)}} \mathbf{a}_k^H \mathbf{v} \} + 2\mathrm{Re}\{ y_k^\star \sqrt{\alpha_k^{(t)}} b_k \}$. Crucially, since $y^\star$ is now fixed, the direct channel term $2\mathrm{Re}\{ y_k^\star \sqrt{\alpha_k^{(t)}} b_k \}$ becomes a constant with respect to $\mathbf{v}$, and thus can be safely omitted from the objective function. Therefore, the equivalent objective for optimizing $\mathbf{v}$ is to maximize:
\begin{equation}
	\sum_{k=1}^{K} \left( 2\mathrm{Re}\left\{ y_k^\star \sqrt{\alpha_k^{(t)}} \mathbf{a}_k^H \mathbf{v} \right\} - |y_k|^2 (\mathbf{v}^H \mathbf{Q}_k \mathbf{v} + \sigma_{total}^2) \right).
\end{equation}

First, we aggregate the terms related to $\mathbf{v}$. Let $\mathbf{c} = \sum_{k} y_k \sqrt{\alpha_k^{(t)}} \mathbf{a}_k$ and $\mathbf{M} = \sum_{k} |y_k|^2 \mathbf{Q}_k$. The maximization problem is equivalent to minimizing its negation:
\begin{equation}
	\min_{\mathbf{v}} \quad \mathbf{v}^H \mathbf{M} \mathbf{v} - 2\mathrm{Re}\{ \mathbf{c}^H \mathbf{v} \}.
\end{equation}
The constant noise term is omitted. Since problem \eqref{eq:obj_30a} is a constrained convex optimization problem. We introduce a Lagrange multiplier $\lambda \ge 0$ to address the total power constraint. The Lagrangian function $\mathcal{L}$ is formulated as:
\begin{equation}
	\mathcal{L}(\mathbf{v}, \lambda) = (\mathbf{v}^H \mathbf{M} \mathbf{v} - 2\mathrm{Re}\{ \mathbf{c}^H \mathbf{v} \}) + \lambda (\mathbf{v}^H \mathbf{Q} \mathbf{v} - P_{max}^{RIS}).
\end{equation}

By rearranging and grouping like terms, the function is rewritten as:
\begin{equation} \label{eq:lagrangian_grouped}
	\mathcal{L}(\mathbf{v}, \lambda) = \mathbf{v}^H (\mathbf{M} + \lambda \mathbf{Q}) \mathbf{v} - 2\mathrm{Re}\{ \mathbf{c}^H \mathbf{v} \} - \lambda P_{max}^{RIS}.
\end{equation}

Since the noise introduced by the active RIS elements is independent, both $\mathbf{M}$ and $\mathbf{Q}$ are diagonal matrices. This implies that the optimization variables corresponding to different RIS elements $v_i$ are mutually independent. Consequently, the problem can be decoupled to solve for each $v_i$ individually.

The matrix form is expanded into the scalar form for each RIS element $i$:
\begin{equation}
	\mathcal{L}_i(v_i) = ([\mathbf{M}]_{ii} + \lambda [\mathbf{Q}]_{ii})|v_i|^2 - 2\mathrm{Re}\{c_i^* v_i\}.
\end{equation}

Taking the partial derivative with respect to $v_i^*$ and setting it to 0:
\begin{equation}
	\frac{\partial \mathcal{L}_i}{\partial v_i^*} = ([\mathbf{M}]_{ii} + \lambda [\mathbf{Q}]_{ii})v_i - c_i = 0.
\end{equation}

Solving for the unconstrained optimal solution $\tilde{v}_i$:
\begin{equation} \label{eq43}
	\tilde{v}_i = \frac{c_i}{[\mathbf{M}]_{ii} + \lambda [\mathbf{Q}]_{ii}}.
\end{equation}

To ensure both the maximum amplification gain constraint \eqref{const:amp} and the individual element saturation constraint \eqref{const:sat} are strictly satisfied, we define an adaptive effective maximum gain amplitude $A_{max, i}^{eff}$ for the $i$-th active element as:
	\begin{equation} \label{eq_eff_amp}
		A_{max, i}^{eff} = \min \left( A_{max}, \sqrt{\frac{P_{sat}}{[\mathbf{Q}]_{ii}}} \right).
	\end{equation}
	If the magnitude of $\tilde{v}_i$ exceeds $A_{max, i}^{eff}$, the solution is projected onto the boundary, i.e., maintaining the phase while clipping the amplitude to $A_{max, i}^{eff}$. The final optimal solution is given by:
\begin{equation} \label{eq44}
	v_i^\star(\lambda) = \begin{cases}
		\tilde{v}_i, & \text{if } |\tilde{v}_i| \le A_{max, i}^{eff}, \\
		A_{max, i}^{eff} \frac{\tilde{v}_i}{|\tilde{v}_i|}, & \text{otherwise}.
	\end{cases}
\end{equation}
At this stage, $\lambda$ is still unknown. We need to find a $\lambda$ such that the total power is exactly equal to or less than $P_{max}^{RIS}$. We define the power residual function with respect to $\lambda$:
\begin{equation} \label{eq:g_lambda}
	g(\lambda) = \sum_{i} [\mathbf{Q}]_{ii} |v_i^\star(\lambda)|^2 - P_{max}^{RIS} = 0.
\end{equation}
Our goal is to solve this equation $g(\lambda) = 0$.

Consider the saturation case where $|v_i^\star| = A_{max}$. In this case, the derivative of $g(\lambda)$ with respect to $\lambda$ is 0. For the non-saturation case, we have:
\begin{equation}
	|v_i^\star(\lambda)|^2 = \frac{|c_i|^2}{([\mathbf{M}]_{ii} + \lambda [\mathbf{Q}]_{ii})^2}.
\end{equation}
Taking the derivative with respect to $\lambda$:
\begin{equation}
	\frac{d}{d\lambda} \left( \frac{|c_i|^2}{([\mathbf{M}]_{ii} + \lambda [\mathbf{Q}]_{ii})^2} \right) = \frac{-2|c_i|^2 \cdot [\mathbf{Q}]_{ii}}{([\mathbf{M}]_{ii} + \lambda [\mathbf{Q}]_{ii})^3}.
\end{equation}
The final expression for the derivative is derived as:
\begin{equation}
	g'(\lambda) = \sum_{i \in \mathcal{U}} [\mathbf{Q}]_{ii} \cdot \left( \frac{-2 [\mathbf{Q}]_{ii} |c_i|^2}{([\mathbf{M}]_{ii} + \lambda [\mathbf{Q}]_{ii})^3} \right),
\end{equation}
where $\mathcal{U}$ denotes the set of RIS elements that are not clipped by the amplitude constraint.

Based on Newton's Iterative method, we perform the following iterative process to quickly search for the optimal value of $\lambda$. The update rule is given by:
\begin{equation} \label{eq49}
	\lambda_{\text{new}} = \lambda_{\text{old}} - \frac{g(\lambda)}{g'(\lambda)}.
\end{equation}
This iteration continues until $g(\lambda) \approx 0$.

Upon obtaining the optimized active reflection coefficient vector $\mathbf{v}^*(\lambda)$, the BS receive beamforming vectors $\{\mathbf{w}_k\}$ are subsequently updated according to \eqref{eq:wk_mrc} to match the newly reconstructed effective channel $\mathbf{h}_{eff,k}(\mathbf{v}^*)$, thereby completing the BCD update cycle for this active phase optimization block.

\section{Position Optimization and Semantic Symbol Length Optimization}
\label{sec:position_opt}  
\subsection{Position Optimization} \label{4A}
With active RIS reflection coefficient vector $\mathbf{v}$ and semantic symbol length set $\mathbf{q}$ fixed, original mixed-integer non-convex optimization problem reduces to a sub-problem with respect to the continuous RIS element position variable $\mathbf{U}$. Due to the fact that the position variables are coupled in channel phase and active noise power in a highly non-linear manner, this sub-problem is non-convex. We propose PGA algorithm combined with Penalty  Method (PM) to address this multi-modal optimization problem.

In this step, the semantic symbol length set $\mathbf{q}$ and the reflection beamformer $\mathbf{v}$ are constants. Our goal is to optimize $\mathbf{U}$ to maximize system performance under the premise of satisfying geometric constraints and the maximum power constraint.

The sub-optimization problem \eqref{obj:sum_sse} can be formulated as:
\begin{subequations} \label{prob:sub_opt_all}
	\begin{align}
		\max_{\mathbf{U}} \quad & f(\mathbf{U}) = \sum_{k=1}^{K} SSE_k (\gamma_k(\mathbf{U}), q_k) \label{eq:sub_obj} \\
		\textrm{s.t.} \quad & x_{min} \le [\mathbf{U}]_{m,n} \le x_{max}, \quad \forall m, n \label{eq:sub_c1} \\
		& [\mathbf{U}]_{m,n+1} - [\mathbf{U}]_{m,n} \ge D_{min}, \quad \forall m, n \label{eq:sub_c2} \\
		& \sum_{k=1}^{K} P_{k} \|\mathrm{diag}(\mathbf{h}_{sr,k}(\mathbf{U}))\mathbf{v}\|_2^2 + \sigma_R^2 \|\mathbf{v}\|_2^2 \le P_{max}^{RIS}, \label{eq:sub_c3} \\
		& P_{out, i}(\mathbf{U}) \le P_{sat}, \quad \forall i \in \{1, \dots, MN\}, \label{eq:sub_c4}
	\end{align}
\end{subequations}
where \eqref{eq:sub_c1} and \eqref{eq:sub_c2} define the set of linear geometric constraints for row-level movement, denoted as $\mathcal{S}_U$. \eqref{eq:sub_c3} represents the non-convex total power constraint specific to the active RIS, and \eqref{eq:sub_c4} restricts the individual amplifier output power to prevent saturation, where $P_{out, i}(\mathbf{U}) = |v_i|^2 \left( \sum_{k=1}^{K} P_k |[\mathbf{h}_{sr,k}(\mathbf{U})]_i|^2 + \sigma_R^2 \right)$. Note that the total power constraint in \eqref{eq:sub_c3} is mathematically equivalent to the sum of the individual output powers over all elements, i.e., $\sum_{i=1}^{MN} P_{out, i}(\mathbf{U}) \le P_{max}^{RIS}$, which explicitly depends on the element positions $\mathbf{U}$ via the first-hop vehicle-to-RIS channel $\mathbf{h}_{sr,k}(\mathbf{U})$.

Although the individual element saturation constraints \eqref{const:sat} (equivalent to \eqref{eq:sub_c4}) theoretically restrict the RIS positions $\mathbf{U}$, their violation is primarily governed by the active reflection coefficients $\mathbf{v}$ rather than the physical positions. Specifically, any minor individual power violations caused by the spatial gradient steps in Step 1 will be strictly and adaptively eliminated by the active phase clipping (with $A_{max, i}^{eff}$) in the subsequent Step 2 of the AO loop. Therefore, to avoid the prohibitive computational overhead of calculating gradients for $MN$ individual non-convex constraints, we only penalize the total power constraint \eqref{const:power} (equivalent to \eqref{eq:sub_c3}) in the augmented loss function $\mathcal{L}(\mathbf{U})$.

The aforementioned sub-problem is mathematically a difficult non-convex optimization problem. Its complexity mainly stems from the following two aspects:

\begin{itemize}
	\item High-frequency oscillation: In high-frequency band, the signal wavelength $\lambda$ is extremely short. The channel response contains the complex exponential term $e^{-j\frac{2\pi}{\lambda}\|\mathbf{u}-\mathbf{p}\|}$. A tiny perturbation in position $\mathbf{U}$ (e.g., $\lambda/2$) can lead to a phase flip of $\pi$, causing objective function to exhibit severe sinusoidal oscillations. This results in extremely dense local optima in the solution space.
	
	\item Non-convex power constraint: The active noise power term $\|\mathbf{H}_{rd}(\mathbf{U})\mathrm{diag}(\mathbf{v})\|_F^2$, when expanded, contains inner product terms of the channels from multiple RIS elements to the base station. This implies that $P_{total}(\mathbf{U})$ itself is a non-convex function with respect to $\mathbf{U}$. Consequently, the feasible region defined by constraint \eqref{eq:sub_c3} may be discontinuous or non-convex, making traditional convex optimization methods directly inapplicable.
\end{itemize}

To address the intractable non-convex power constraint \eqref{eq:sub_c3}, we employ the PM, transforming the hard constraint into a soft penalty term in the objective function.

We first define the power violation function $g(\mathbf{U})$ as:
\begin{equation}
	g(\mathbf{U}) = \sum_{k=1}^{K} P_{k} \|\mathrm{diag}(\mathbf{h}_{sr,k}(\mathbf{U}))\mathbf{v}\|_2^2 + \sigma_R^2 \|\mathbf{v}\|_2^2 - P_{max}^{RIS}. \label{eq:power_violation}
\end{equation}
We construct an augmented loss function $\mathcal{L}(\mathbf{U})$, aiming to minimize the negative SSE along with a power penalty term:
\begin{equation}
	\mathcal{L}(\mathbf{U}) = \underbrace{-\sum_{k=1}^{K} SSE_k(\mathbf{U})}_{\text{Performance Metric}} + \underbrace{\rho \cdot [\max(0, g(\mathbf{U}))]^2}_{\text{Quadratic Penalty Term}}, \label{eq:augmented_loss}
\end{equation}
where $\rho > 0$ is the penalty factor. When the total power constraint in \eqref{eq:sub_c3} is satisfied, i.e., $g(\mathbf{U}) \le 0$, the penalty term is 0, and the optimization process is driven entirely by SSE. Conversely, when the constraint is violated, i.e., $g(\mathbf{U}) > 0$, the penalty term introduces a large positive gradient, forcing the search direction to rapidly return towards the feasible region that satisfies the power constraint.

At this point, the original constrained problem \eqref{eq:sub_obj} transforms into the following partially unconstrained problem defined on a convex set:
\begin{equation}
	\min_{\mathbf{U} \in \mathcal{S}_U} \mathcal{L}(\mathbf{U}).
\end{equation}

To solve the transformed problem, we employ the Projected Gradient Descent (PGD) method. This method consists of two alternating steps: 
\subsubsection{Gradient Descent}
Leveraging Automatic Differentiation techniques, we apply the chain rule directly to the complex cascaded channel and active noise terms to compute the gradient of the loss function with respect to each position coordinate $x_{m,n}$, denoted as $\nabla_{\mathbf{U}}\mathcal{L}$.

In the $t$-th iteration, we update the position along the direction of the negative gradient:
\begin{equation} \label{eq54}
	\tilde{\mathbf{U}}^{(t+1)} = \mathbf{U}^{(t)} - \eta \cdot \nabla_{\mathbf{U}}\mathcal{L}(\mathbf{U}^{(t)}),
\end{equation}
where $\eta$ is the learning rate, and $\tilde{\mathbf{U}}^{(t+1)}$ represents the intermediate position variable that has not yet satisfied the geometric constraints.
\subsubsection{Geometric Projection}
We project the intermediate variable $\tilde{\mathbf{U}}^{(t+1)}$ back onto the linear geometric constraint set $\mathcal{S}_U$. Since the row-level movements are decoupled among different rows, the projection operator $\mathcal{P}_{\mathcal{S}_U}$ can be decomposed into independent operations for each row $m$.

To rigorously address the ordered box constraints, which comprise both the physical boundaries and the minimum inter-element spacing, we employ an exact mathematical projection based on the Pool Adjacent Violators Algorithm (PAVA). The exact projection for the intermediate coordinates $\tilde{\mathbf{x}} = [\tilde{\mathbf{U}}^{(t+1)}]_{m,:}$ in the $m$-th row is executed through the following sequence:
\begin{itemize}
	\item \textbf{Coordinate Transformation:} We first convert the coupled minimum spacing constraints $x_{n+1} - x_n \ge D_{min}$ into standard monotonicity constraints. By defining an auxiliary sequence $y_n$ for $n \in \{1, \dots, N\}$, the spatial non-collision requirement is equivalently transformed into a monotonic non-decreasing sequence condition:
	\begin{equation} \label{eq56}
			y_n = \tilde{x}_n - (n-1)D_{min}, \quad \text{s.t.} \quad y_1 \le y_2 \le \dots \le y_N.
	\end{equation}
	\item \textbf{Isotonic Regression (PAVA):} We apply PAVA to the sequence $\mathbf{y}$ to obtain a monotonically non-decreasing sequence $\mathbf{z}$. Specifically, PAVA initializes $\mathbf{z} = \mathbf{y}$ and scans the elements sequentially. Whenever a monotonicity violation is detected (i.e., $z_n > z_{n+1}$), the algorithm merges the violating adjacent elements into a single block and replaces their values with their arithmetic mean:
		\begin{equation} \label{57}
			z_n = z_{n+1} = \frac{z_n + z_{n+1}}{2}.
		\end{equation}
	If this newly updated block violates the order with its preceding block, they are further merged and averaged. This backward pooling process repeats until the entire sequence strictly satisfies the non-decreasing property, yielding the exact Euclidean projection of $\mathbf{y}$ onto the monotonic constraint set.
	\item \textbf{Box Clamping:} To enforce the guide rail boundary constraints $\mathcal{P}$, we rigidly clip the sequence $\mathbf{z}$. Given that $\mathbf{z}$ is already monotonic, applying a global clipping preserves the monotonic property without causing secondary constraint violations:
	\begin{equation} \label{58}
			z'_n = \min \left( \max \left( z_n, x_{min} \right), x_{max} - (N-1)D_{min} \right).
	\end{equation}
	\item \textbf{Coordinate Recovery:} Finally, we perform the inverse transformation to recover the updated feasible positions for the RIS elements:
	\begin{equation} \label{59}
			[\mathbf{U}^{(t+1)}]_{m,n} = z'_n + (n-1)D_{min}.
	\end{equation}
\end{itemize}

This PAVA-based ordered box projection theoretically guarantees to return the exact Euclidean projection, strictly ensuring that the final updated position $\mathbf{U}^{(t+1)}$ completely resides within the feasible convex set $\mathcal{S}_U$ without any boundary violations or spacing conflicts.
\subsection{Semantic Signal Length Optimization} \label{4B}

After completing the alternating iterations for the active beamforming $\mathbf{v}$ and the RIS element positions $\mathbf{U}$, the physical layer state of the system (i.e., channel gains and interference environment) is fully determined. At this stage, the sub-problem regarding the semantic symbol length set $\mathbf{q} = \{q_1, \dots, q_K\}$ exhibits a unique mathematical structure. Addressing the discrete integer characteristics of this variable and its unimodal property within the objective function, we employ an efficient one-dimensional discrete greedy search algorithm for the solution.

With $\mathbf{U}$ and $\mathbf{v}$ fixed, the received SINR $\gamma_k$ for each vehicle $k$ becomes a constant. Upon observing the original optimization problem \eqref{obj:sum_sse}, we find:

\begin{enumerate}
	\item Objective Function Separability: Sum-SSE is simply linear superposition of SSE of each individual vehicle.
	\item Constraint Decoupling: The semantic similarity constraint \eqref{const:qos} is solely related to the vehicle's own $q_k$ and $\gamma_k$. Furthermore, $q_k$ does not involve the active RIS power constraint (which is determined solely by physical layer variables).
\end{enumerate}

Consequently, the original multi-dimensional combinatorial optimization problem \eqref{obj:sum_sse} can be decomposed into $K$ mutually independent single-variable integer programming problems. For any arbitrary vehicle $k$, optimization objective is:

\begin{subequations} \label{prob:semantic_single_all}
	\begin{align}
		\max_{q_k} \quad & g_k(q_k) = \frac{I}{L \cdot q_k} \cdot \xi_k(\gamma_k, q_k), \label{eq:sem_obj} \\
		\textrm{s.t.} \quad & \xi_k(\gamma_k, q_k) \ge \xi_{th} \label{eq:sem_c1}, \\
		& q_k \in \{1, 2, \dots, Q_{max}\} \label{eq:sem_c2}.
	\end{align}
\end{subequations}

Since the SINR is held constant, the semantic similarity transforms into a function solely dependent on $q_k$. Generally, the semantic similarity exhibits a monotonically increasing trend with the increase of $q_k$. This implies the existence of a unique optimal $q_k^*$ that achieves the optimal trade-off between semantic accuracy and resource consumption. Given the finite discrete value range of $q_k$, we adopt a 1D Greedy Search strategy. This approach guarantees finding the global optimal solution within the discrete domain with extremely low computational complexity.

For each vehicle $k$, the following steps are executed in parallel:
\begin{enumerate}
	\item Feasible Region Determination:
	Based on the currently fixed SINR $\gamma_k$, we first identify the set of candidate coding lengths that satisfy constraint \eqref{eq:sem_c1}:
	\begin{equation} \label{eq58}
		\mathcal{K}_{feasible} = \{ q_k \mid q_k \in [1, Q_{max}], \; \xi_k(\gamma_k, q_k) \ge \xi_{th} \}.
	\end{equation}
	
	If $\mathcal{K}_{feasible}$ is empty, it implies that the current physical channel quality is insufficient to support valid semantic transmission. In this case, we set $q_k$ to the maximum allowable value.
	
	\item Exhaustive Search:
	We iterate through every candidate $q_k$ within the feasible set $\mathcal{K}_{feasible}$ and calculate the corresponding SSE according to \eqref{eq:sse_def}. Since the range $Q_{max}$ is typically small, the computational overhead of this exhaustive search process is negligible.
	
	\item Greedy Selection:
	We select the $q_k$ value that maximizes $SSE_k(q_k)$ as the updated value for the current iteration:
	\begin{equation} \label{eq59}
		q_k^{(t+1)} = \arg \max_{q_k \in \mathcal{K}_{feasible}} SSE_k(q_k).
	\end{equation}
\end{enumerate}

By nesting the aforementioned Position Optimization using PGA and Semantic Signal Length Optimization using Greedy Search within the AO framework, the proposed algorithm can rapidly converge to a suboptimal solution of the joint system.

\begin{algorithm}[!t]
	\caption{Chronological Two-Timescale Joint Optimization and Adaptation Algorithm}
	\label{alg:two_timescale}
	\small
	\begin{algorithmic}[1]
		\REQUIRE Predicted CSI $\hat{\mathbf{h}}(t)$ at slot $t$, frame length $N_{slots}$, total slots $T_{total}$, initial variables.
		\ENSURE Optimized sequences $\{\mathbf{q}^*(t), \mathbf{v}^*(t), \mathbf{U}^*(t)\}$ for all $t \in \{1, \dots, T_{total}\}$.
		
		\FOR{each time slot $t = 1, \dots, T_{total}$}
		\STATE Acquire newly predicted CSI $\hat{\mathbf{h}}(t)$ at current slot $t$.
		\STATE \textbf{Initialize:} AO index $i=0$, random $\mathbf{v}^{(0)}$, initial $\mathbf{q}^{(0)}$, and $\mathbf{U}^{(0)} \leftarrow \mathbf{U}^*(t-1)$ (if $t > 1$).
		\REPEAT
		
		\STATE \textit{// Step 1: Position Optimization (Only at Frame Boundaries)}
		\IF{$(t-1) \pmod{N_{slots}} == 0$}
		\STATE Initialize $\mathbf{U}_{best} \leftarrow \mathbf{U}^{(i)}$, $\mathcal{O}_{best} \leftarrow f(\mathbf{v}^{(i)}, \mathbf{U}^{(i)}, \mathbf{q}^{(i)})$.
		\FOR{inner iteration $iter = 1, \dots, T_{pos}$}
		\STATE Compute gradient $\nabla_{\mathbf{U}}\mathcal{L}(\mathbf{U})$.
		\STATE Update intermediate $\tilde{\mathbf{U}}$ via \eqref{eq54}. 
		\STATE Project to feasible $\mathbf{U}^{(i, iter)}$ via PAVA \eqref{eq56},\eqref{57},\eqref{58},\eqref{59}.
		\IF{$f(\mathbf{v}^{(i)}, \mathbf{U}^{(i, iter)}, \mathbf{q}^{(i)}) > \mathcal{O}_{best}$}
		\STATE $\mathbf{U}_{best} \leftarrow \mathbf{U}^{(i, iter)}$, $\mathcal{O}_{best} \leftarrow f(\mathbf{v}^{(i)}, \mathbf{U}^{(i, iter)}, \mathbf{q}^{(i)})$.
		\ENDIF
		\ENDFOR
		\STATE Retain best feasible positions: $\mathbf{U}^{(i+1)} \leftarrow \mathbf{U}_{best}$.
		\ELSE
		\STATE Fix positions to previous slot: $\mathbf{U}^{(i+1)} \leftarrow \mathbf{U}^*(t-1)$.
		\ENDIF
		
		\STATE \textit{// Step 2: Active Phase Optimization (Section~\ref{sec:phase_opt})}
		\STATE Calculate semantic weights $\alpha_k$.
		\STATE Update auxiliary $y_k$ via \eqref{eq35}. 
		\STATE Calculate unconstrained $\tilde{\mathbf{v}}$ via \eqref{eq43}.
		\STATE Find optimal $\lambda$ using Newton's method via \eqref{eq49}.
		\STATE Project $\tilde{\mathbf{v}}$ onto constraint to get $\mathbf{v}^{(i+1)}$ via \eqref{eq44}.
		\STATE Update $\mathbf{w}_k^{(i+1)}$ based on $\mathbf{v}^{(i+1)}$ via \eqref{eq:wk_mrc}.
		
		\STATE \textit{// Step 3: Semantic Symbol Length Optimization (Section~\ref{4B})}
		\FOR{each vehicle $k = 1, \dots, K$}
		\STATE Identify feasible set $\mathcal{K}_{feasible}$ via \eqref{eq58}.
		\STATE Update $q_k^{(i+1)}$ via 1D Greedy Search \eqref{eq59}.
		\ENDFOR
		
		\STATE Update outer iteration count $i \leftarrow i + 1$.
		\UNTIL{Sum-SSE converges or Max Iterations reached.}
		\STATE $\mathbf{U}^*(t) \leftarrow \mathbf{U}^{(i)}$, $\mathbf{v}^*(t) \leftarrow \mathbf{v}^{(i)}$, $\mathbf{q}^*(t) \leftarrow \mathbf{q}^{(i)}$.
		
		\IF{$(t-1) \pmod{N_{slots}} == 0$}
		\STATE \textbf{Hardware Execution:} Mechanically deploy RIS elements to $\mathbf{U}^*(t)$.
		\ENDIF
		
		\ENDFOR
		
		\RETURN Optimal sequences $\{\mathbf{q}^*(t), \mathbf{v}^*(t), \mathbf{U}^*(t)\}, \forall t$.
	\end{algorithmic}
\end{algorithm}

\subsection{Two-Timescale Joint Optimization Framework}
\label{sec:two_timescale_framework}
	
To bridge the gap between theoretical gains and hardware execution constraints, we integrate the decoupled sub-problems into a unified, chronological two-timescale framework. Instead of treating the large and small timescales as isolated processes, the system operates slot-by-slot over a duration of $T_{total}$ slots, where a large time frame $T_{large}$ comprises $N_{slots}$ small time slots. 
	
At any given time slot $t$, the optimization behavior is determined by whether the slot marks the beginning of a new frame $T_{large}$ (i.e., $(t-1) \pmod{N_{slots}} = 0$):
\begin{itemize}
	\item \textbf{Joint Optimization at Frame Boundaries:} When $t$ is at a frame boundary, the BS executes the complete joint AO loop using the predicted CSI $\hat{\mathbf{h}}(t)$ to optimize the positions $\mathbf{U}$ (Step 1), phases $\mathbf{v}$ (Step 2), and semantic parameters $\mathbf{q}$ (Step 3). The RIS elements are then mechanically deployed to the optimized locations $\mathbf{U}^*(t)$ and locked.
		
	\item \textbf{Fast Adaptation Within Frames:} For any intermediate slot $t$ within the frame, the positions are fixed to those of the previous slot ($\mathbf{U}^*(t) = \mathbf{U}^*(t-1)$) to bypass mechanical latency. The BS only runs a low-complexity adaptation loop to update the electronic phases $\mathbf{v}$ and semantic lengths $\mathbf{q}$ based on the newly predicted instantaneous CSI $\hat{\mathbf{h}}(t)$.
\end{itemize}
This chronological operation protocol, detailed in Algorithm \ref{alg:two_timescale}, guarantees that mechanical movements occur only at coarse boundaries, while electronic and semantic parameters remain agile.

\subsection{Convergence Analysis}
\label{sec:convergence}
In this subsection, we rigorously analyze the convergence behavior of the proposed AO algorithm utilized in the large-timescale joint optimization phase. Let $\mathcal{O}(\mathbf{v}, \mathbf{U}, \mathbf{q})$ denote the objective function (Sum-SSE) of problem \eqref{obj:sum_sse}. 

\textbf{Theorem 1.} \textit{The proposed joint optimization algorithm guarantees the monotonic convergence of the system objective value (Sum-SSE).}

\textit{Proof:} At the $t$-th iteration, the algorithm sequentially updates the coupled variables:

1) \textit{Phase Optimization:} With fixed $\mathbf{U}^{(t)}$ and $\mathbf{q}^{(t)}$, the SCA method constructs a surrogate lower bound based on the continuously differentiable fitted spline. Furthermore, the Quadratic Transform yields an exact reformulation for the fractional SINR. According to the minorize-maximization (MM) framework, updating $\mathbf{v}^{(t+1)}$ by exactly solving the reformulated convex sub-problem guarantees that the original objective is non-decreasing:
\begin{equation}
	\mathcal{O}(\mathbf{v}^{(t+1)}, \mathbf{U}^{(t)}, \mathbf{q}^{(t)}) \ge \mathcal{O}(\mathbf{v}^{(t)}, \mathbf{U}^{(t)}, \mathbf{q}^{(t)}).
\end{equation}

2) \textit{Position Optimization:} With fixed $\mathbf{v}^{(t+1)}$ and $\mathbf{q}^{(t)}$, the PGA algorithm explores the continuous position space. Finding an exact global optimum or a strict stationary point for this highly non-convex sub-problem within finite gradient steps is mathematically intractable. Therefore, our PGA incorporates a best-so-far tracking strategy. By explicitly evaluating the projected feasible positions across the $T_{pos}$ inner iterations, the algorithm retains the updated position $\mathbf{U}^{(t+1)}$ that maximizes the Sum-SSE. In the worst case, if no gradient step yields an improvement, it strictly maintains $\mathbf{U}^{(t)}$. This finite-step heuristic inherently guarantees that the objective value never decreases:
\begin{equation}
	\mathcal{O}(\mathbf{v}^{(t+1)}, \mathbf{U}^{(t+1)}, \mathbf{q}^{(t)}) \ge \mathcal{O}(\mathbf{v}^{(t+1)}, \mathbf{U}^{(t)}, \mathbf{q}^{(t)}).
\end{equation}

3) \textit{Semantic Length Optimization:} With fixed $\mathbf{v}^{(t+1)}$ and $\mathbf{U}^{(t+1)}$, the 1D greedy search explicitly evaluates all discrete integer coding lengths within the feasible set $\mathcal{K}_{feasible}$. Since this step performs an exact maximization in the 1D discrete domain, it guarantees:
\begin{equation}
	\mathcal{O}(\mathbf{v}^{(t+1)}, \mathbf{U}^{(t+1)}, \mathbf{q}^{(t+1)}) \ge \mathcal{O}(\mathbf{v}^{(t+1)}, \mathbf{U}^{(t+1)}, \mathbf{q}^{(t)}).
\end{equation}

Combining the above inequalities, the overall objective function value is monotonically non-decreasing over successive AO iterations. Furthermore, the system Sum-SSE is strictly upper-bounded by the finite transmission bandwidth, the bounded semantic similarity $\xi_k \in [0,1]$, and the hardware power constraints (maximum amplification gain $A_{max}$ and total power $P_{max}^{RIS}$). 

According to the Monotone Convergence Theorem, a strictly bounded and monotonically non-decreasing sequence $\mathcal{O}^{(t)}$ is mathematically guaranteed to converge to a finite value. While claiming rigorous convergence to a KKT stationary point is relaxed due to the finite-step exploration nature of the MINLP position sub-problem, this monotonic improvement ensures algorithm stability. The highly competitive performance and superiority of this converged sub-optimal solution will be extensively validated in the numerical results. $\hfill \blacksquare$

\subsection{Computational Complexity Analysis}
This subsection analyzes the computational complexity of the proposed joint optimization algorithm, as summarized in Algorithm \ref{alg:two_timescale}. The overall complexity is primarily dominated by the three steps within the AO framework. Let $I_{AO}$ denote the number of AO iterations. The complexity of each step is derived as follows:

\subsubsection{Active Phase Optimization}
In this step, the dominant operations involve the calculation of auxiliary variables and the update of reflection coefficients. 
Calculating the effective cascaded channel involves matrix–vector multiplications between the RIS–BS channel and the beamforming vector, resulting in a complexity of $\mathcal{O}(K N_{BS} N_{RIS})$, where $N_{RIS} = MN$ is the total number of RIS elements.
Since the matrices $\mathbf{M}$ and $\mathbf{Q}$ in \eqref{eq43} are diagonal, the calculation of the unconstrained solution $\tilde{\mathbf{v}}$ and the Newton's method for $\lambda$ requires only linear complexity $\mathcal{O}(N_{RIS})$.
Thus, the total complexity of this step is $\mathcal{O}(K N_{BS} N_{RIS} + I_{New}N_{RIS})$, where $I_{New}$ is the number of Newton iterations.

\subsubsection{Position Optimization}
The complexity of the PGA-based position optimization is dominated by the gradient calculation. 
Using the chain rule, computing the gradient of the loss function $\nabla_{\mathbf{U}}\mathcal{L}$ with respect to the positions requires traversing all $K$ vehicles and calculating the derivatives of the channel vectors, which scales linearly with the number of antennas and RIS elements.
Therefore, the complexity per gradient descent step is $\mathcal{O}(K N_{BS} N_{RIS})$.
Given $T_{pos}$ iterations for the PGA process, the complexity of this step is $\mathcal{O}(T_{pos} K N_{BS} N_{RIS})$.

\subsubsection{Semantic Signal Length Optimization}
The 1D greedy search for the semantic symbol length $q_k$ is performed independently for each vehicle. 
For each vehicle, we search through a maximum of $Q_{max}$ candidate values. The calculation of SSE is a scalar operation with $\mathcal{O}(1)$ cost.
Consequently, the complexity of this step is $\mathcal{O}(K Q_{max})$.

\subsubsection{Total Complexity}
Based on the above analysis, the overall computational complexity of the proposed Algorithm \ref{alg:two_timescale} is:
\begin{equation}
	\mathcal{O}\left(I_{AO} \left( (1 + T_{pos}) K N_{BS} N_{RIS} + K Q_{max} \right) \right).
\end{equation}
It is worth noting that the proposed algorithm avoids high-complexity matrix inversion operations, ensuring its scalability in large-scale RIS deployments. In contrast, the heuristic Quantum-behaved Particle Swarm Optimization (QPSO) algorithm \cite{zhou2025movable,ref6} benchmark entails a complexity of $\mathcal{O}(T_{QPSO} P_{size} N_{RIS}^2)$, which suffers from the curse of dimensionality as the number of particles $P_{size}$ and iterations $T_{QPSO}$ must increase exponentially to maintain performance in high-dimensional spaces.

\begin{table}[htbp]
	\centering
	\caption{Per-Iteration Execution Time of Each Subproblem}
	\label{tab:runtime_comparison}
	\begin{tabular}{lc}
		\hline
		\textbf{Subproblem} & \textbf{Time (s)} \\ \hline
		Proposed PGA Position Update (one iteration) & 0.424 \\
		Proposed SCA Phase Update (one iteration) & 0.027 \\
		Proposed Greedy Semantic Length Selection & 0.010 \\
		APS Position Search (one iteration) & 1.485 \\
		QPSO Heuristic (full search, no AO) & 8.575 \\ \hline
	\end{tabular}
\end{table}

To quantitatively evaluate the computational overhead and verify the engineering feasibility of our algorithm, we measured the average per-iteration execution time of each subproblem on an Intel Core i7-14650HX CPU (sequential execution in a pure Python 3.8 environment). As detailed in Table~\ref{tab:runtime_comparison}, the closed-form phase-only update and semantic length selection are extremely lightweight, requiring merely 0.027 s and 0.010 s, respectively. For position optimization, although AD introduces gradient computation overhead, the proposed PGA (0.424 s) leverages exact descent directions to significantly outperform the coordinate-descent APS baseline (1.485 s). In contrast, the derivative-free QPSO heuristic demands up to 8.575 s for its blind search without alternating optimization (AO). This confirms that AD effectively accelerates convergence rather than hindering scalability.

Furthermore, this AO framework fits seamlessly into our practical two-timescale scheme. On the large-timescale where macroscopic topology varies slowly, a full joint optimization typically converges in 3 AO iterations, taking approximately 1.384 s ($3 \times (0.424 + 0.027 + 0.010)$ s). Executing this at frame boundaries is highly feasible, as the vehicle displacement during this latency (e.g., $\sim$27.6 m at highway speeds) remains well within the spatial correlation distance of large-scale fading. Conversely, on the small-timescale within frames, physical positions are locked. The base station only performs phase and semantic updates, requiring just 37 ms ($0.027 + 0.010$ s) per iteration to swiftly track fast fading. Finally, while these benchmarks reflect interpreted Python execution, practical industrial deployments utilizing C++, parallelized $K$-vehicle gradient computations, and edge GPU acceleration will further compress latencies to the sub-millisecond level, fully satisfying ultra-low latency constraints.

\begin{table}[t]
	\centering
	\caption{Simulation Parameters}
	\label{tab:sim_params}
	\renewcommand{\arraystretch}{1.3} 
	\begin{tabular}{|l|c|} 
		\hline
		\textbf{Parameter} & \textbf{Value} \\
		\hline
		BS Coordinates & $[125.0, -30.0, 25.0]$ m \\
		\hline
		RIS Center Coordinates & $[175.0, 20.0, 15.0]$ m \\
		\hline
		Path Loss Exponent (Direct Link $\alpha_{d}$) & 3.5 \\
		\hline
		Path Loss Exponent (RIS-assisted Link $\alpha_{sr}$) & 2.2 \\
		\hline
		BS Antenna Gain ($G_{bs}$) & 8 dB \\
		\hline
		Vehicle Antenna Gain ($G_{veh}$) & 3 dB \\
		\hline
		Number of RIS Elements ($M \times N$) & $5 \times 5$ \\
		\hline
		Active Reflection Gain & 20 dB \\
		\hline
		Maximum Transmit Power ($P_{k}$) & 23 dBm \\
		\hline
		Number of Vehicles ($K$) & 20 \\
		\hline
		Average Vehicle Velocity & 20 m/s \\
		\hline
		Velocity Standard Deviation ($\sigma_{v}$) & 3.0 m/s \\
		\hline
		Avg. Semantic information per Sentence ($I$) & 100.0 sut/sentence \\
		\hline
		Avg. Sentence Length ($L$) & 10.0 words/sentence \\
		\hline
		Semantic Similarity Threshold ($\xi_{th}$) & 0.9 \\
		\hline
		AO Iterations ($I_{AO}$) & 3 \\
		\hline
		Newton's Method Iterations ($I_{New}$) & 5 \\
		\hline
		Gradient Descent Iterations ($T_{pos}$) & 20 \\
		\hline
		Gradient Descent Learning Rate & 0.0005 \\
		\hline
		random seeds & 112, 163, 480\\
		\hline
	\end{tabular}
\end{table}
\section{Simulation Results}
\label{sec:simulation}    
In this section, simulations are conducted using Python 3.8 to validate the effectiveness of the proposed method. The detailed parameter settings are summarized in Table \ref{tab:sim_params}. Specifically, all numerical results are averaged over 3 fixed random seeds and 50 large-timescale frames, where 3 small-timescale slots within each large-timescale frame are sampled via Monte Carlo snapshots to evaluate the phase and semantic adaptation results. We compare the performance of the following seven different schemes:

\begin{enumerate}
	\item Full AO: Based on the proposed method, performing alternating optimization for the RIS element positions $\mathbf{U}$, active reflection coefficients $\mathbf{v}$, and semantic signal length $\mathbf{q}$.
	
	\item Phase only: Optimizing only the reflection coefficients $\mathbf{v}$ and semantic signal length $\mathbf{q}$, while keeping the RIS element positions fixed.
	
	\item Position only: Optimizing only the RIS element positions $\mathbf{U}$ and semantic signal length $\mathbf{q}$, without optimizing the reflection phases.
	
	\item QPSO: Jointly optimizing positions, phases, and $\mathbf{q}$ using the QPSO algorithm \cite{zhou2025movable}, serving as a benchmark comparison for the proposed solution.
	
	\item APS position: Given discrete hardware locations satisfying the constraints, the optimal discrete positions are obtained using a one-dimensional alternating search method, while the optimal semantic symbol length is greedily searched.
	
	\item Random position: The positions of the RIS elements are randomly reset, and the optimal semantic symbol length is greedily searched, serving as a baseline comparison for position optimization.
	
	\item Passive RIS: Full AO based on passive RIS architecture. Unlike active RIS, passive RIS introduces neither thermal noise nor additional signal amplification gain. This serves as a baseline to verify the necessity of active RIS in vehicular environments, particularly under long-distance transmission conditions.
\end{enumerate}
\begin{figure*}[!t]
	\centering
	\begin{minipage}[t]{0.32\textwidth}
		\centering
		\includegraphics[width=\linewidth]{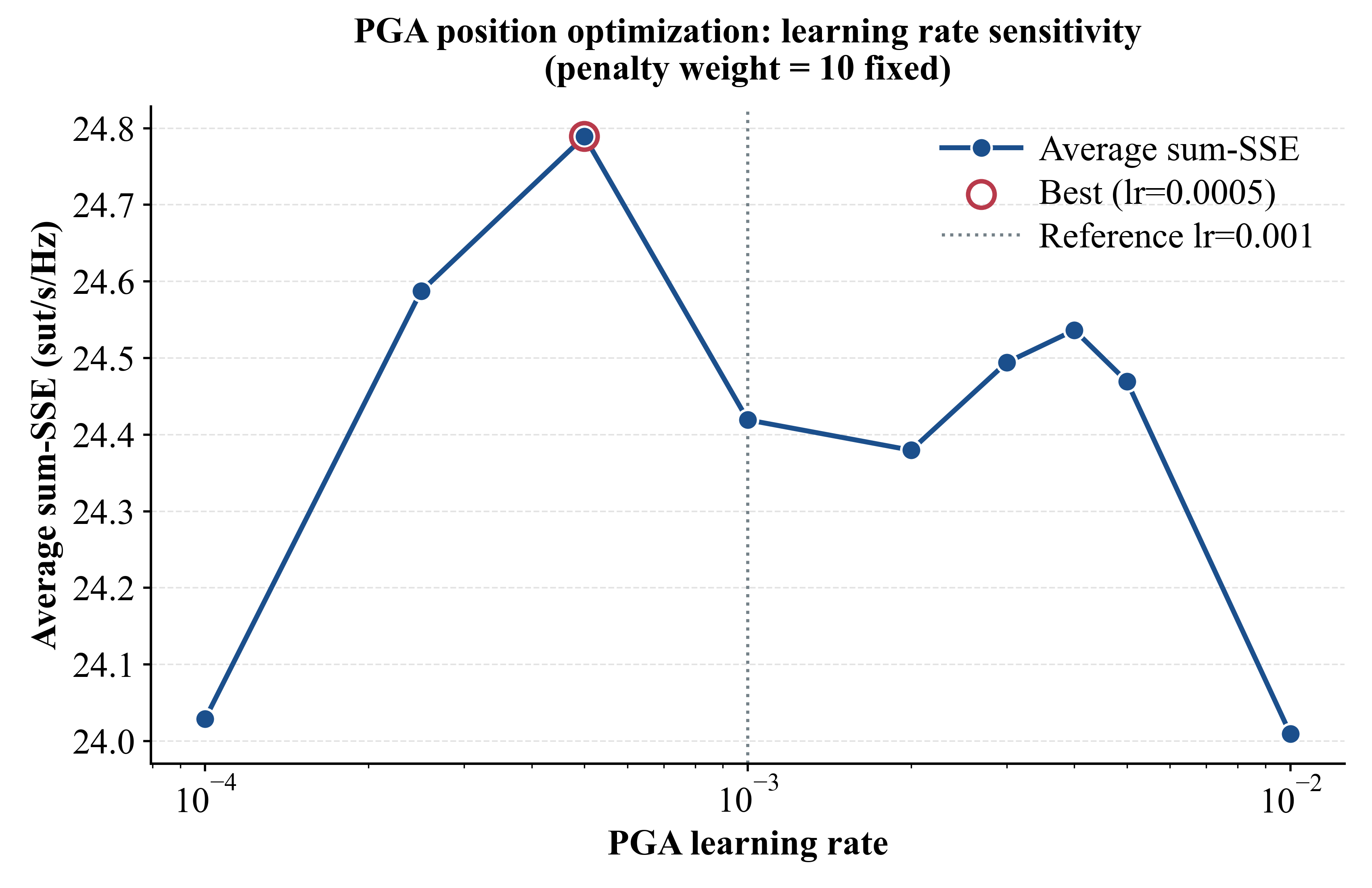}
		\caption{Learning rate sensitivity of PGA.}
		\label{fig_99}
	\end{minipage}
	\hfill 
	\begin{minipage}[t]{0.32\textwidth}
		\centering
		\includegraphics[width=\linewidth]{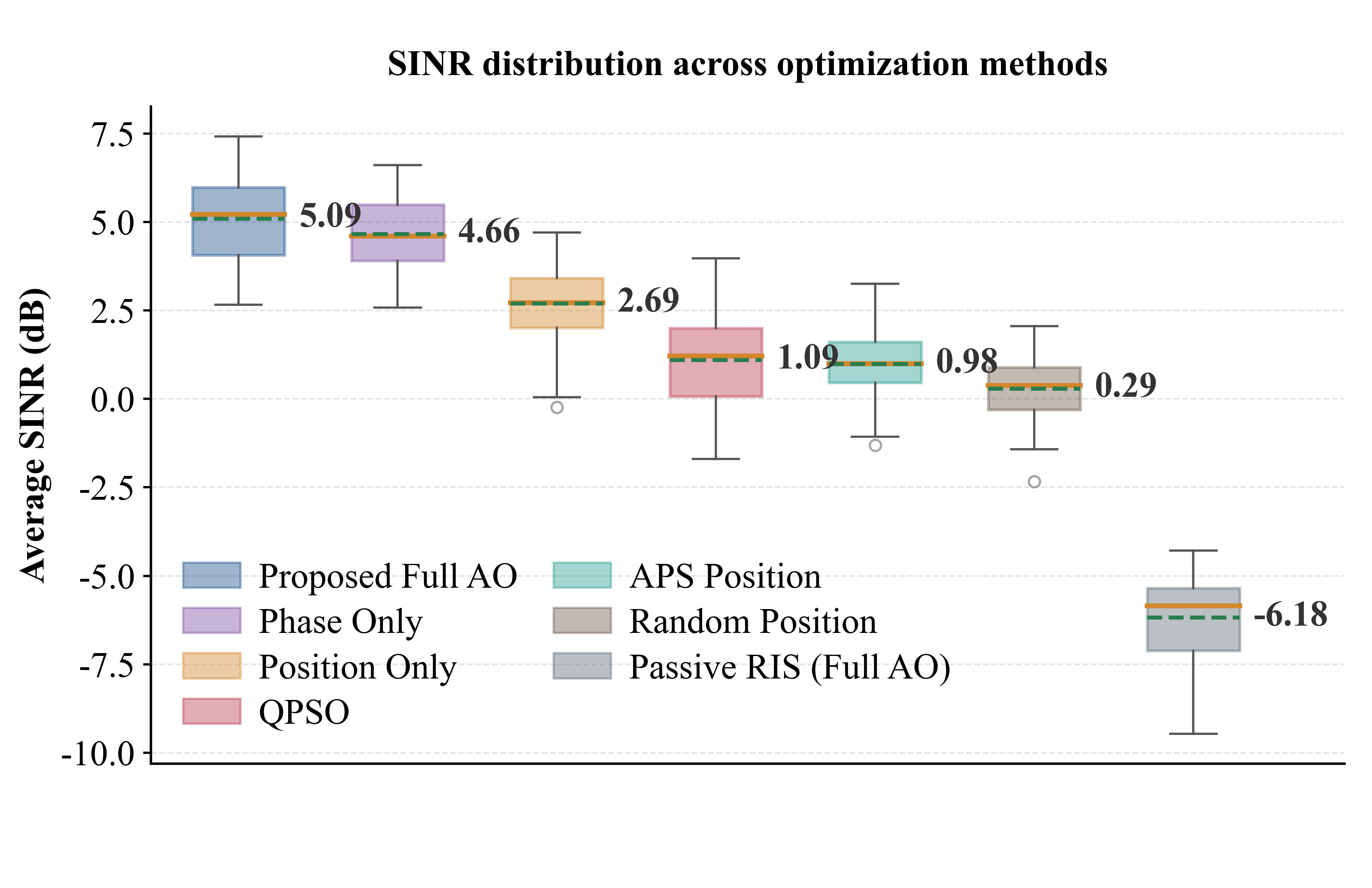}
		\caption{Average SINR for different methods}
		\label{fig_2}
	\end{minipage}
	\hfill 
	\begin{minipage}[t]{0.32\textwidth}
		\centering
		\includegraphics[width=\linewidth]{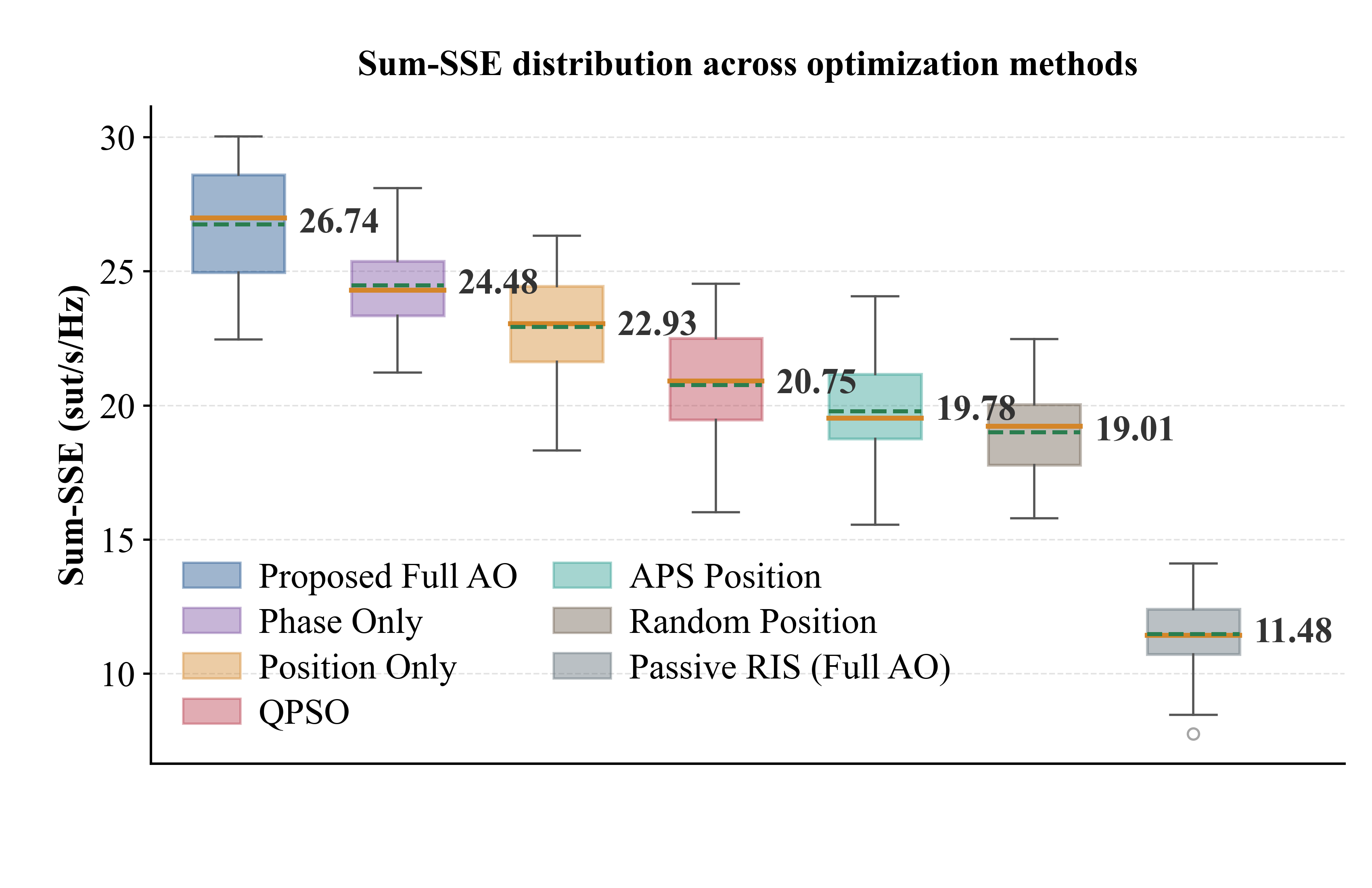}
		\caption{Sum-SSE for different methods}
		\label{fig_3}
	\end{minipage}
\end{figure*}

Fig.~\ref{fig_99} illustrates the performance sensitivity of the proposed PGA-based algorithm under varying learning rates $\eta$. The average Sum-SSE exhibits a bimodal trend, showing performance degradation at both extremely low and high learning rates. Specifically, a small learning rate (e.g., $\eta \le 10^{-4}$) slows down convergence within the finite inner iterations, while an excessively large learning rate (e.g., $\eta \ge 10^{-2}$) causes the coordinate updates to overshoot optimal regions, leading to oscillations and constraint violations. Despite slight fluctuations in the intermediate region due to the non-convex coordinate space, the peak performance is achieved at $\eta = 0.0005$. Therefore, we select $\eta = 0.0005$ as the baseline learning rate to ensure both convergence speed and stable optimization.

Fig. \ref{fig_2} presents the statistical distribution of the achievable SINR over 50 time-slot updates. The proposed Full AO method achieves an average SINR of 5.09 dB, outperforming the ``Phase Only'' scheme (4.66 dB) and significantly surpassing the ``Position Only'' scheme (1.09 dB). This demonstrates that the reflection phase plays a more dominant role in beamforming alignment than element positioning. Furthermore, the proposed Full AO yields a substantial 11.27 dB gain over the ``Passive RIS'' scheme (-6.18 dB), verifying the necessity of active reflection in long-distance V2I environments. In terms of position optimization, our PGA-based algorithm achieves massive gains over the ``APS Position'' (2.69 dB) and ``Random Position'' (0.29 dB) baselines. This is because continuous gradient-based tracking avoids the accumulation of quantization errors inherent in discrete alternating coordinate search. Additionally, due to the tendency of meta-heuristics to get trapped in local optima in high-dimensional search spaces, the QPSO scheme only achieves an average SINR of 0.98 dB.

\begin{figure*}[!t]
	\centering
	\begin{minipage}[t]{0.32\textwidth}
		\centering
		\includegraphics[width=\linewidth]{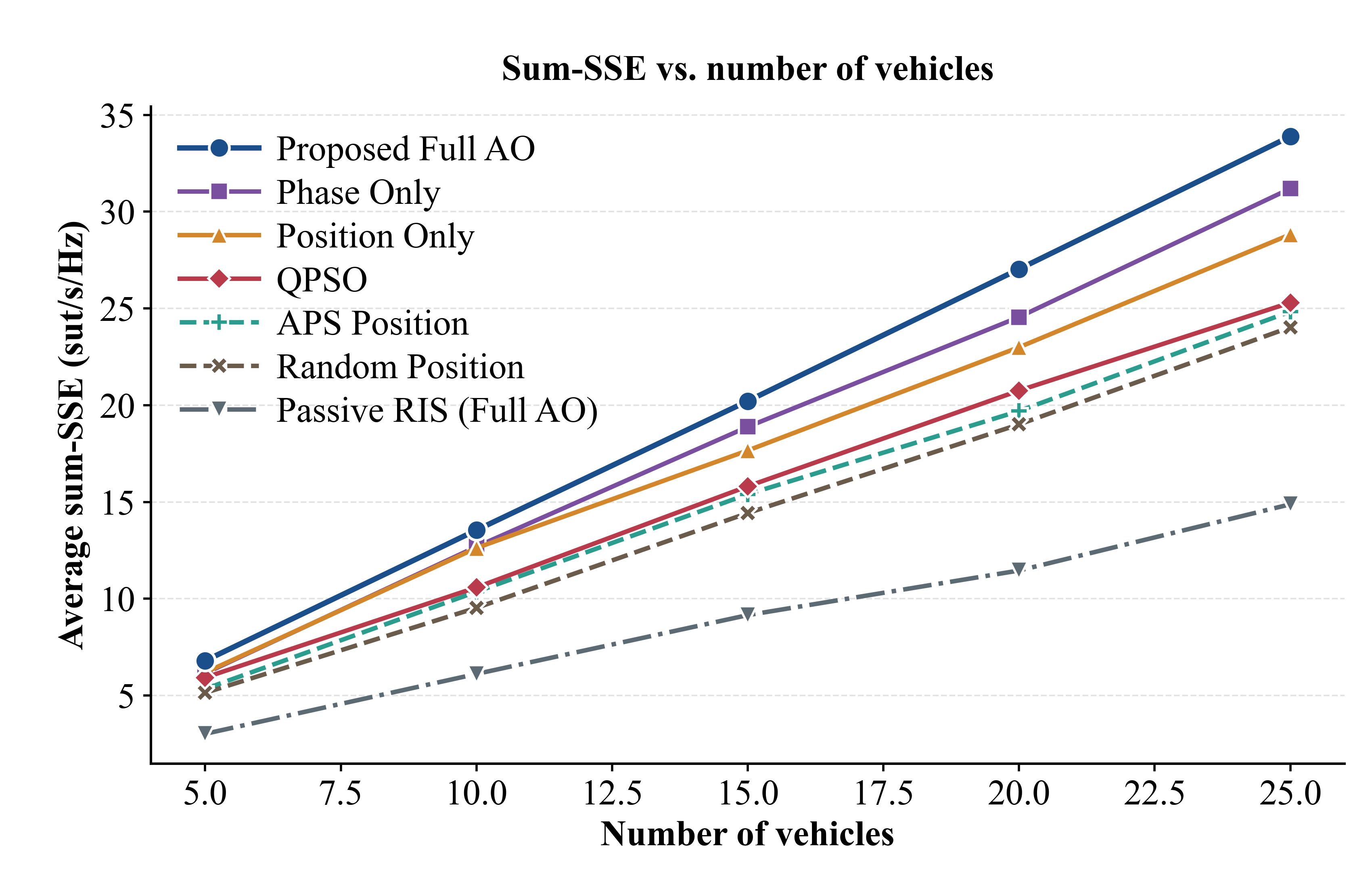}
		\caption{Average Sum-SSE for different number of vehicles.}
		\label{fig_4}
	\end{minipage}
	\hfill 
	\begin{minipage}[t]{0.32\textwidth}
		\centering
		\includegraphics[width=\linewidth]{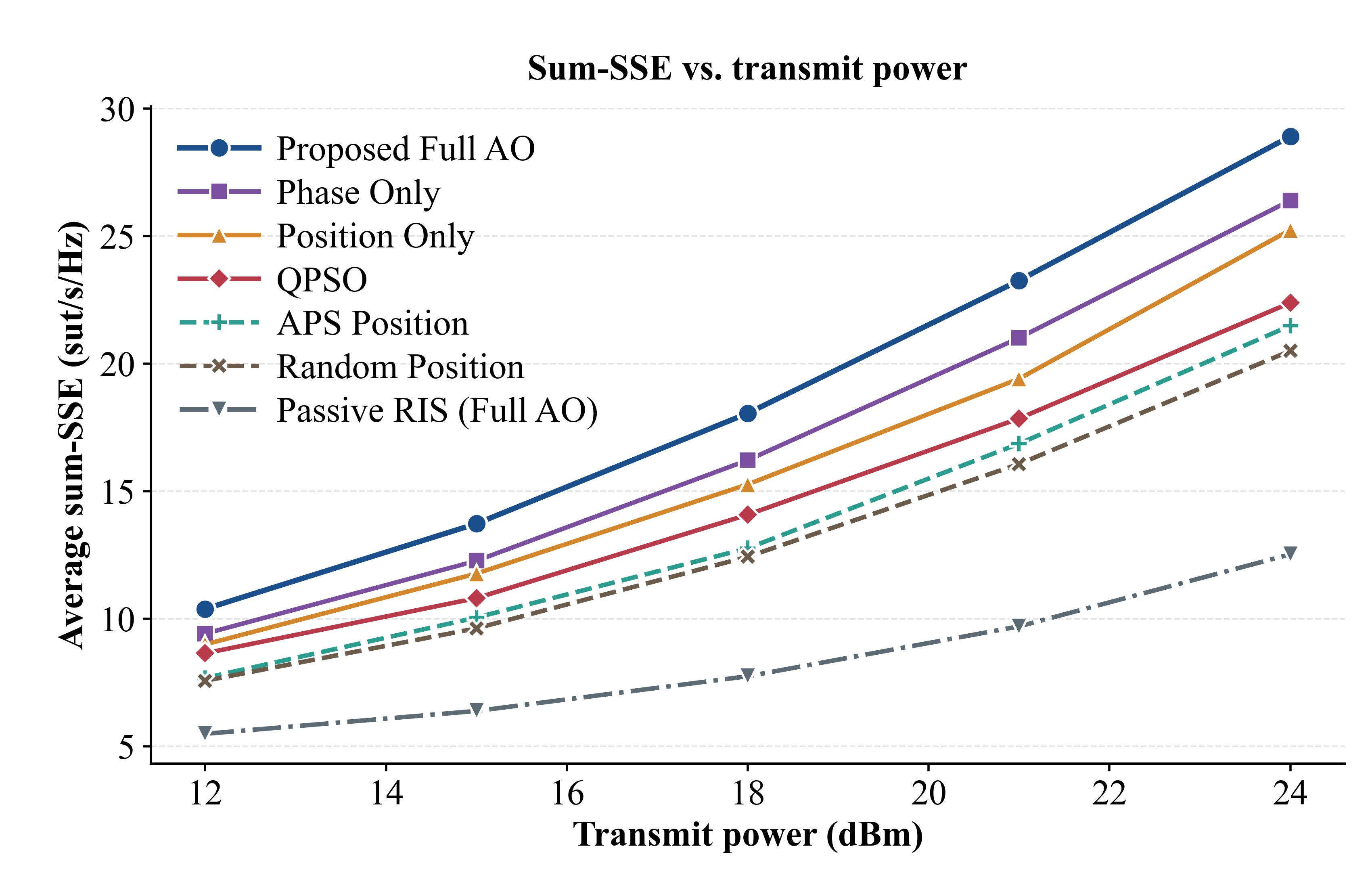}
		\caption{Average Sum-SSE for different transmit power.}
		\label{fig_5}
	\end{minipage}
	\hfill 
	\begin{minipage}[t]{0.32\textwidth}
		\centering
		\includegraphics[width=\linewidth]{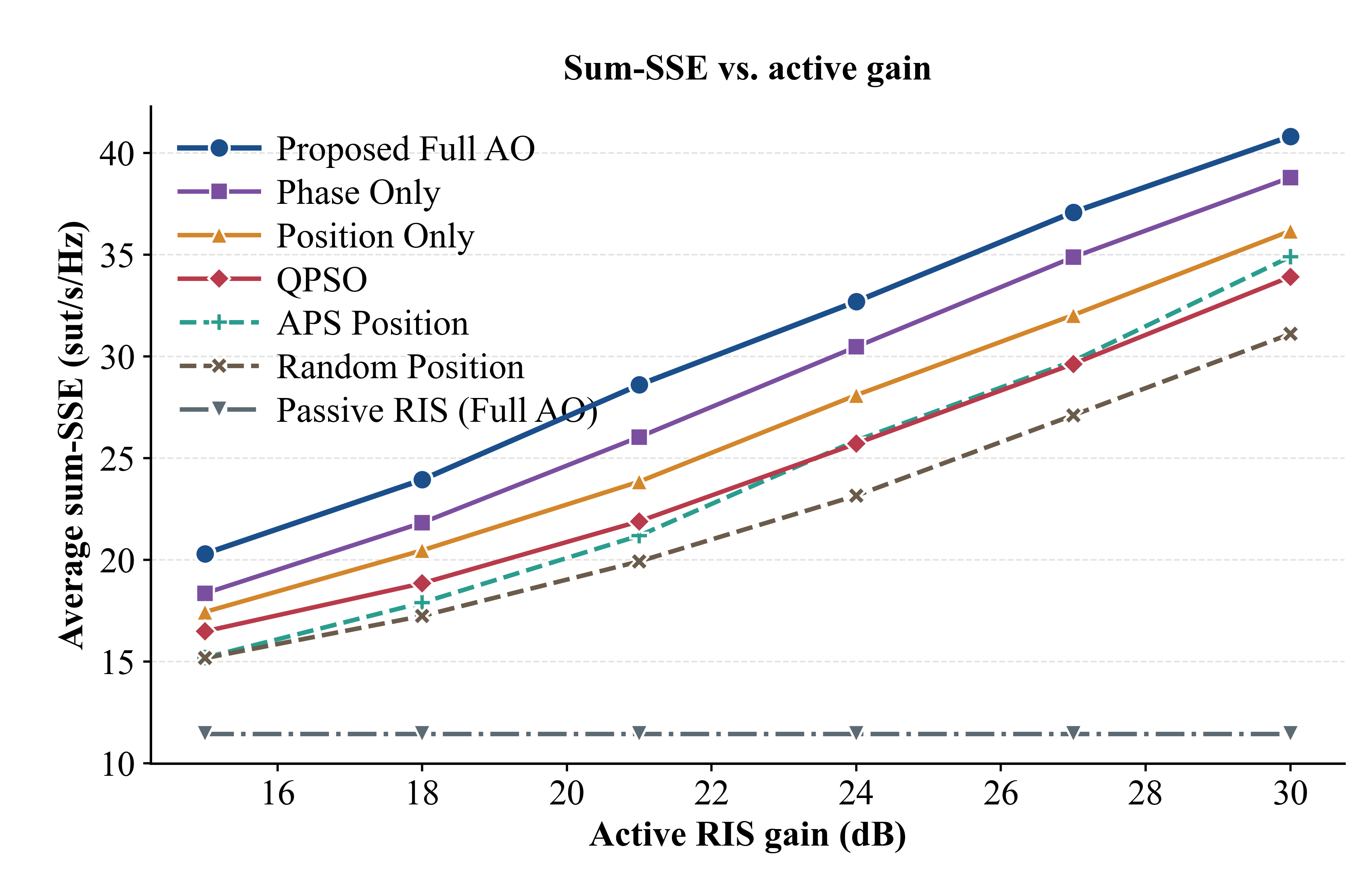}
		\caption{Average Sum-SSE for different active gain.}
		\label{fig_9}
	\end{minipage}
\end{figure*}
\begin{figure*}[!t]
	\centering
	\begin{minipage}[t]{0.32\textwidth}
		\centering
		\includegraphics[width=\linewidth]{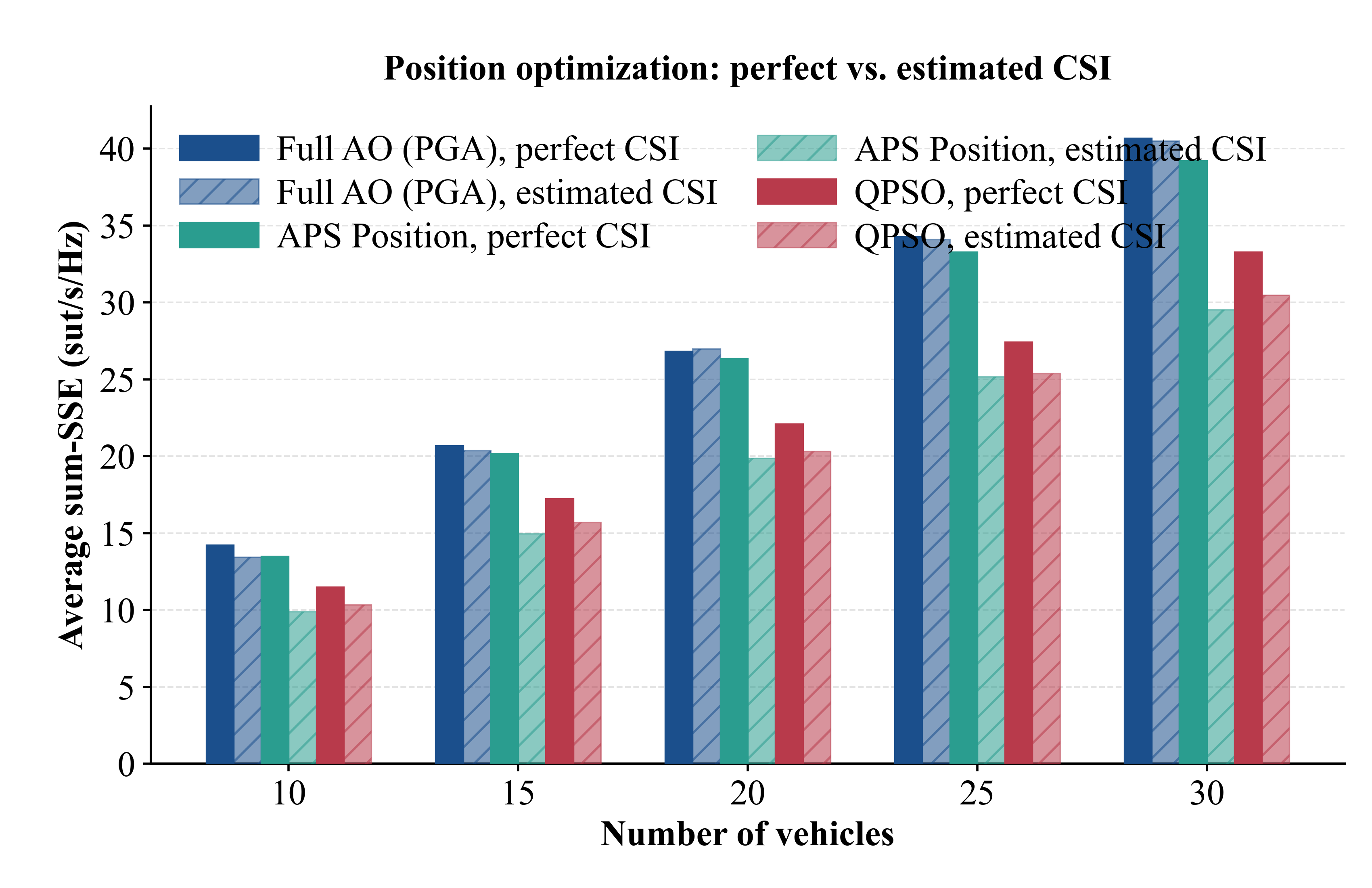}
		\caption{Impact of CSI estimation errors on the Sum-SSE of different optimization schemes.}
		\label{fig_6}
	\end{minipage}
	\hfill 
	\begin{minipage}[t]{0.32\textwidth}
		\centering
		\includegraphics[width=\linewidth]{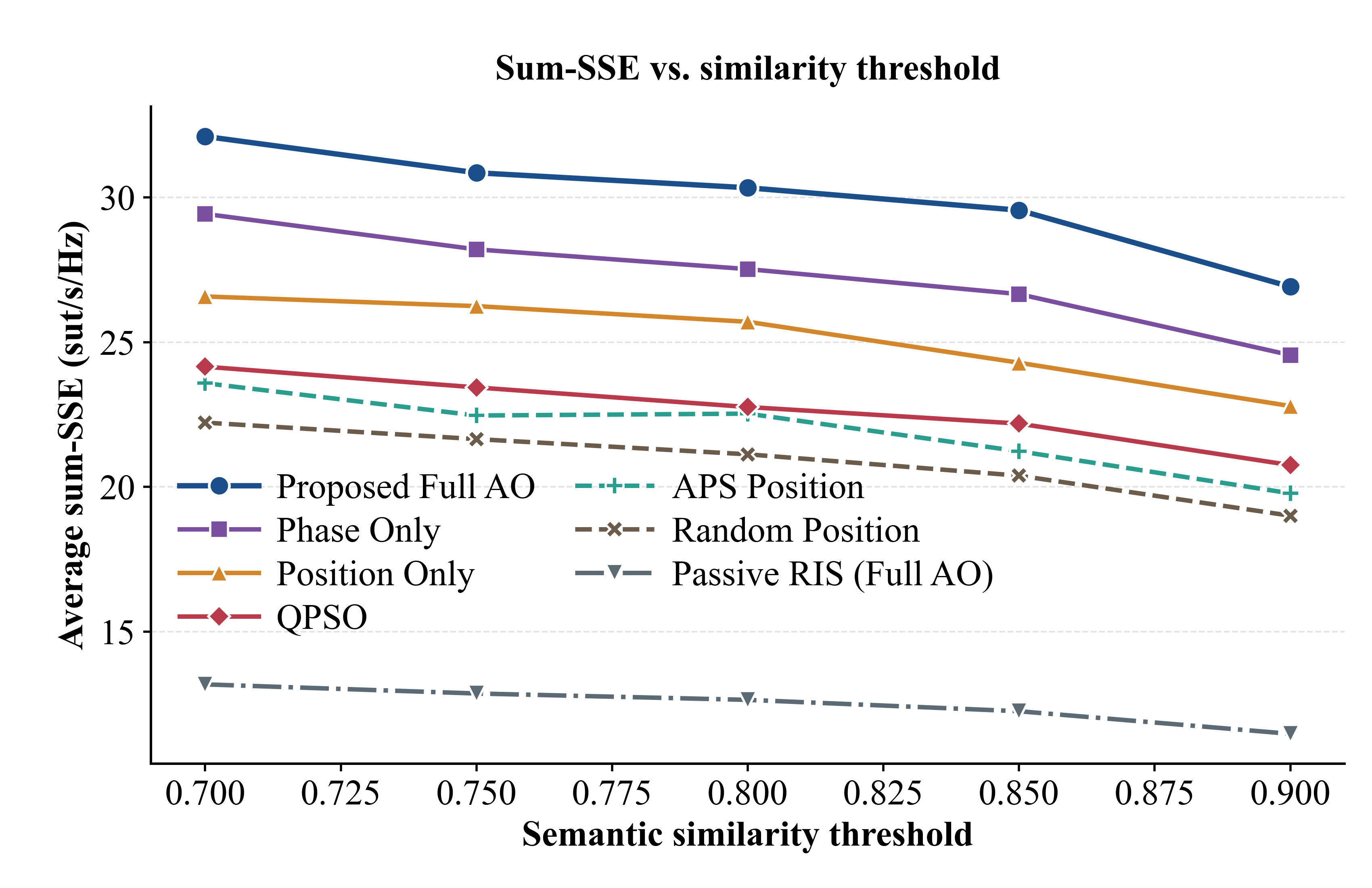}
		\caption{Average Sum-SSE for different semantic similarity threshold}
		\label{fig_7}
	\end{minipage}
	\hfill 
	\begin{minipage}[t]{0.32\textwidth}
		\centering
		\includegraphics[width=\linewidth]{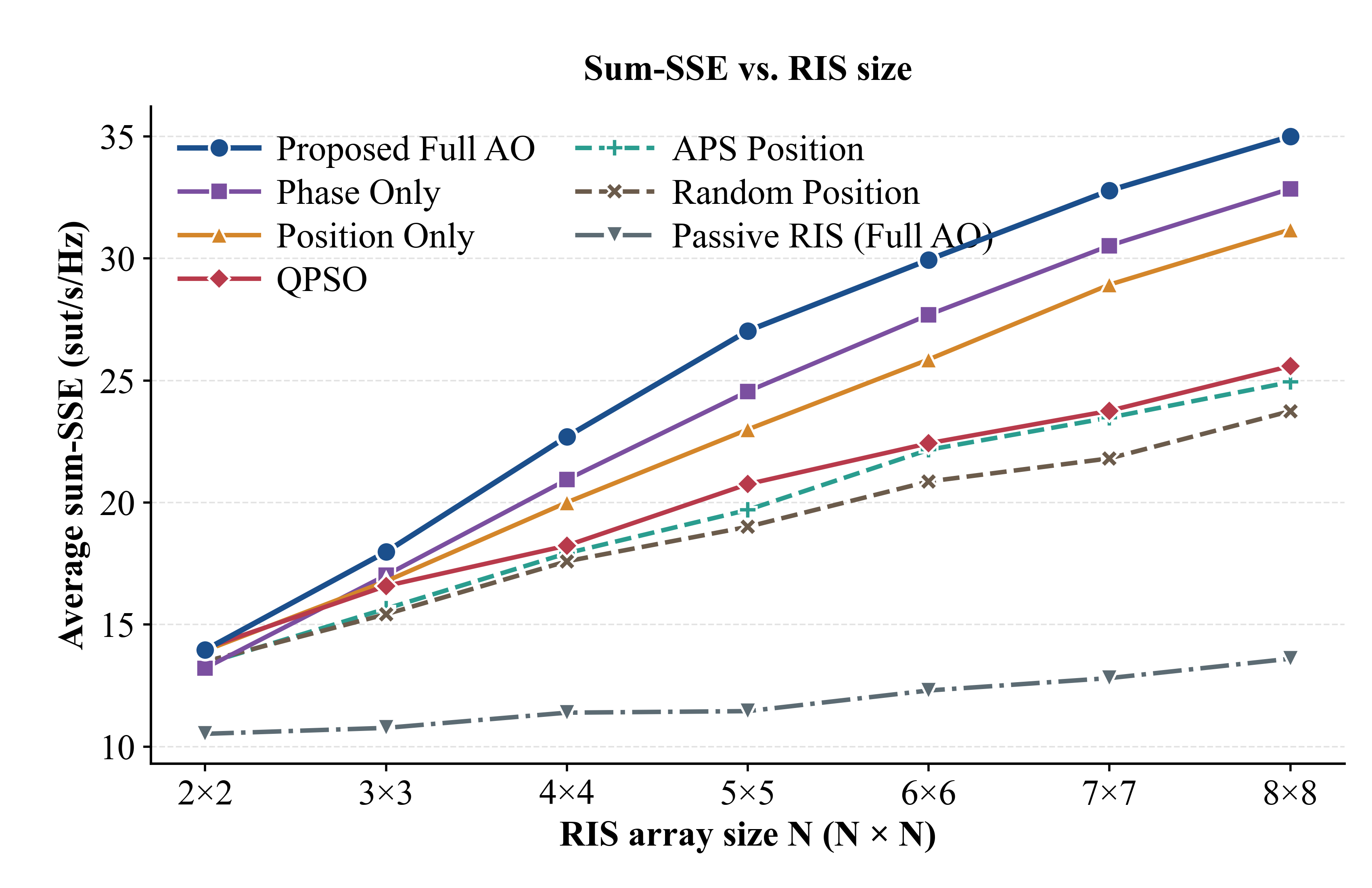}
		\caption{Average Sum-SSE for different RIS size}
		\label{fig_8}
	\end{minipage}
\end{figure*}
Fig. \ref{fig_3} depicts the statistical performance of the system Sum-SSE. The proposed Full AO achieves a superior average Sum-SSE of 26.74 suts/Hz, representing a 132.9\% and 35.2\% improvement over the ``Passive RIS'' (11.48 suts/Hz) and QPSO (19.78 suts/Hz) schemes, respectively. Crucially, the Sum-SSE improvement of Full AO over the ``Phase Only'' scheme (24.48 suts/Hz) is more pronounced (a 9.2\% increase of 2.26 suts/Hz) than its SINR counterpart (only a 0.25 dB increase). This is because position optimization offers higher spatial degrees of freedom to physically bypass deep fading dips on weak links. Since the semantic similarity function is highly sensitive to channel variations in the low-SINR regime, improving the channel quality of weak users yields a much higher marginal SSE gain than further optimizing already strong (high-SINR) users. The ``APS Position'' and ``Random Position'' schemes only achieve 22.93 suts/Hz and 19.01 suts/Hz, respectively, further demonstrating the superior convergence of our PGA algorithm.

Fig.~\ref{fig_4} compares the average Sum-SSE against the number of vehicles $K$. As $K$ increases, the Sum-SSE of all schemes scales up due to the spatial multiplexing gain. Specifically, the proposed Full AO maintains a distinct advantage across all vehicle counts, achieving 33.89~suts/Hz at $K=25$, which represents a 127.8\% and 8.6\% improvement over ``Passive RIS'' (14.88~suts/Hz) and ``Phase Only'' (31.19~suts/Hz), respectively. When the vehicle count is small ($K=5$), the ``Phase Only'' (6.14~suts/Hz) and ``Position Only'' (6.19~suts/Hz) schemes perform comparably because a simple geometry is sufficient for a small number of users. However, as $K$ increases, the environment complexity rises, and ``Position Only'' falls behind ``Phase Only'' (e.g., at $K=25$, ``Position Only'' is only 28.82~suts/Hz compared to 31.19~suts/Hz of ``Phase Only''). This demonstrates that solely adjusting positions cannot adapt to dense, time-varying vehicular distributions without agile phase control. Although QPSO performs comparably to ``APS Position'' at $K=5$, its performance deteriorates significantly in denser networks, achieving only 25.30~suts/Hz at $K=25$. This degradation is caused by the exponential expansion of the search space with $K$, which slows down convergence and severely compromises the solution quality of meta-heuristics.

Fig.~\ref{fig_5} analyzes the average Sum-SSE against varying transmit powers. As the power increases from 12~dBm to 24~dBm, the Sum-SSE of Full AO rises from 10.38~suts/Hz to 28.91~suts/Hz. At 24~dBm, the proposed Full AO achieves a 9.5\% and 130.9\% improvement over the ``Phase Only'' (26.40~suts/Hz) and ``Passive RIS'' (12.52~suts/Hz) schemes, respectively. Furthermore, comparing the three position-only optimization baselines, namely the PGA-based ``Position Only'', the classic ``APS Position'', and the ``Random Position'', reveals that our PGA-based scheme (25.24~suts/Hz at 24~dBm) consistently outperforms the other two. Specifically, ``Position Only'' yields a 17.4\% and 23.1\% performance increase over ``APS Position'' (21.49~suts/Hz) and ``Random Position'' (20.50~suts/Hz), respectively. Under identical fixed-phase conditions, this directly demonstrates that continuous gradient-based position tracking achieves superior channel reconstruction compared to the standard alternating coordinate search and random deployment.

Fig.~\ref{fig_9} illustrates the average Sum-SSE versus active reflection gain. The Sum-SSE of Full AO increases from 20.31~suts/Hz to 40.81~suts/Hz as the gain escalates from 15~dB to 30~dB, representing a 5.2\% and 256.1\% improvement over the ``Phase Only'' (38.79~suts/Hz) and ``Passive RIS'' (11.46~suts/Hz) schemes at 30~dB, respectively. Other active schemes also scale up monotonically. Specifically, among the position-only baselines, the PGA-based ``Position Only'' scheme (reaching 36.17~suts/Hz at 30~dB) outperforms the classic ``APS Position'' (34.90~suts/Hz) by 3.6\% and ``Random Position'' (31.10~suts/Hz) by 16.3\%. This superiority underscores that continuous search via PGA is highly effective for exploiting spatial degrees of freedom, outperforming the standard alternating search even without active phase optimization.

Fig. \ref{fig_6} evaluates the robustness of different optimization schemes against CSI estimation errors under varying vehicle counts. For a fair comparison, the APS baseline is augmented with our proposed phase optimization. Under imperfect CSI, joint optimization is performed using the estimated CSI, while actual performance is evaluated against the true instantaneous CSI. As shown, the Sum-SSE monotonically increases with the vehicle count due to spatial multiplexing. Notably, compared to perfect CSI, the proposed Full AO exhibits superior robustness with less than 6\% performance loss. The QPSO method shows moderate sensitivity with roughly 8\%--10\% loss, whereas the APS method suffers severe degradation (approximately 25\% loss). This is because the sequential coordinate search in APS is highly sensitive to CSI mismatch; position evaluation errors accumulate across search steps, leading to suboptimal deployment.

Fig.~\ref{fig_7} evaluates the system Sum-SSE against semantic similarity thresholds $\xi_{th}$. As the quality constraints tighten from 0.7 to 0.9, all schemes exhibit a monotonic decline in Sum-SSE. Notably, at $\xi_{th} = 0.9$, the proposed Full AO achieves 26.91~suts/Hz, yielding a 9.6\% and 134.6\% improvement over the ``Phase Only'' (24.55~suts/Hz) and ``Passive RIS'' (11.47~suts/Hz) schemes, respectively. Across all thresholds, the PGA-based ``Position Only'' scheme (e.g., 22.79~suts/Hz at $\xi_{th}=0.9$) consistently outperforms the classic ``APS Position'' (19.78~suts/Hz) and ``Random Position'' (19.01~suts/Hz). In particular, at $\xi_{th} = 0.9$, ``Position Only'' outperforms ``APS Position'' by 15.2\% and ``Random Position'' by 19.9\%. This further validates that the continuous geometric flexibility of PGA-based deployment provides more robust channel quality than standard alternating searches or random topologies under various semantic constraints.

Fig.~\ref{fig_8} analyzes the Sum-SSE against the RIS array size $N$ (configured as $N \times N$ arrays). While Passive RIS remains stagnant (13.60~suts/Hz at $8\times8$), active schemes show significant gains. At $8\times8$, the proposed Full AO (35.00~suts/Hz) achieves a 6.5\% and 157.4\% improvement over the ``Phase Only'' (32.85~suts/Hz) and ``Passive RIS'' (13.60~suts/Hz) schemes, respectively. For position optimization comparisons, the PGA-based ``Position Only'' scheme achieves 31.18~suts/Hz at $8\times8$, substantially outperforming the classic ``APS Position'' (24.94~suts/Hz) by 25.0\% and ``Random Position'' (23.75~suts/Hz) by 31.3\%. This gap demonstrates that the continuous, joint-coordinate search achieved by PGA scales more effectively with larger arrays than the decoupled alternating search, which is prone to error accumulation in high-dimensional settings.
\section{Conclusion}
\label{sec:conclusion}    
In this paper, we have investigated a novel RM-A-RIS assisted vehicular semantic communication framework. To address the dual challenges of severe multiplicative path loss and fast-varying channels in IoV environments, we proposed a hardware-efficient row-level movable architecture. By jointly optimizing the RIS element positions, active reflection coefficients, and semantic symbol lengths, we formulated a non-convex problem to maximize the system Sum-SSE. We developed an efficient AO algorithm that integrates SCA-based phase optimization, penalty-based PGA for position updates, and a greedy search for semantic parameters.

Extensive simulation results demonstrate that the proposed RM-A-RIS scheme exhibits distinct advantages over traditional configurations, achieving a Sum-SSE of 26.74 suts/Hz, which represents improvements of 9.2\% and 132.9\% compared to the fixed-position active RIS and conventional passive RIS benchmarks, respectively. Our analysis reveals two critical insights: 1) Active signal amplification is indispensable for overcoming the severe cascaded path loss in vehicular scenarios, yielding a substantial 11.27 dB gain in received SINR compared to the passive RIS system; 2) the optimization of element positions provides additional spatial degrees of freedom that effectively adapt to dynamic vehicular distributions and enhance transmission quality, particularly in large-scale array deployments. Future work will investigate the robust beamforming design considering imperfect CSI for movable elements.



\end{document}